\documentclass[twocolumn,amsmath,amssymb,superscriptaddress]{revtex4}
\pdfoutput=1

\usepackage[pdftex]{graphicx}
\usepackage[pdftex,colorlinks=true,urlcolor=blue,linkcolor=black]{hyperref} 

\usepackage{graphicx}
\usepackage{dcolumn}
\usepackage{bm}
\usepackage{datetime}
\usepackage{color}
\usepackage{units}

\usepackage{enumitem}
\setitemize{leftmargin=*} 

\usepackage{ulem}				

\newcommand{\ket}[1]{\vert #1 \rangle}	

\begin{document}

\title{Observation of Pair Condensation in the Quasi-2D BEC-BCS Crossover}

\author{M. G. Ries\footnote{To whom correspondence should be addressed. E-mail: \href{mailto:mries@physi.uni-heidelberg.de}{mries@physi.uni-heidelberg.de}}}
\thanks{These authors contributed equally to this work.}
\affiliation{Physikalisches Institut, Ruprecht-Karls-Universit\"at Heidelberg, 69120 Heidelberg, Germany}

\author{A. N. Wenz}
\thanks{These authors contributed equally to this work.}
\affiliation{Physikalisches Institut, Ruprecht-Karls-Universit\"at Heidelberg, 69120 Heidelberg, Germany}

\author{G. Z\"urn}
\altaffiliation{These authors contributed equally to this work.}
\affiliation{Physikalisches Institut, Ruprecht-Karls-Universit\"at Heidelberg, 69120 Heidelberg, Germany}

\author{L. Bayha}
\affiliation{Physikalisches Institut, Ruprecht-Karls-Universit\"at Heidelberg, 69120 Heidelberg, Germany}

\author{I. Boettcher}
\affiliation{Institut f\"ur Theoretische Physik, Ruprecht-Karls-Universit\"at Heidelberg, 69120 Heidelberg, Germany}

\author{D. Kedar}
\affiliation{Physikalisches Institut, Ruprecht-Karls-Universit\"at Heidelberg, 69120 Heidelberg, Germany}

\author{P. A. Murthy}
\affiliation{Physikalisches Institut, Ruprecht-Karls-Universit\"at Heidelberg, 69120 Heidelberg, Germany}

\author{M. Neidig}
\affiliation{Physikalisches Institut, Ruprecht-Karls-Universit\"at Heidelberg, 69120 Heidelberg, Germany}

\author{T. Lompe}
\altaffiliation[Present address: ]{MIT-Harvard Center for Ultracold Atoms, MIT, Cambridge, MA 02139, USA.}
\affiliation{Physikalisches Institut, Ruprecht-Karls-Universit\"at Heidelberg, 69120 Heidelberg, Germany}

\author{S. Jochim}
\affiliation{Physikalisches Institut, Ruprecht-Karls-Universit\"at Heidelberg, 69120 Heidelberg, Germany}

\date{\today}

\begin{abstract}
The condensation of fermion pairs lies at the heart of superfluidity. However, for strongly correlated systems with reduced dimensionality the mechanisms of pairing and condensation are still not fully understood. 
In our experiment we use ultracold atoms as a generic model system to study the phase transition from a normal to a condensed phase in a strongly interacting quasi-two-dimensional Fermi gas. Using a novel method, we obtain the in situ pair momentum distribution of the strongly interacting system and observe the emergence of a low-momentum condensate at low temperatures. By tuning temperature and interaction strength we map out the phase diagram of the quasi-2D BEC-BCS crossover.
\end{abstract}

\maketitle

The characteristics of quantum many-body systems are strongly affected by their dimensionality and the strength of interparticle correlations. 
In particular, strongly correlated two-dimensional fermionic systems have been of interest because of their connection to high-$T_c$ superconductivity. 
Although they have been the subject of intense theoretical studies \cite{Norman2011,Lee2006,Levinsen2014,Randeria1989,Iskin2009,Bertaina2011,Bauer2014,Matsumoto2014}, a complete theoretical framework has not yet been established. 

Ultracold quantum gases are an ideal realization for exploring strongly interacting 2D Fermi gases, as they offer the possibility of independently tuning the dimensionality and the strength of interparticle interactions.
Reducing the dimensionality \cite{Bloch2008} led to the observation of a Berezinskii-Kosterlitz-Thouless (BKT) type phase transition to a superfluid phase in weakly interacting 2D Bose gases \cite{Hadzibabic2006,Desbuquois2012}. Tuning the strength of interactions in a three-dimensional two-component Fermi gas made it possible to explore the crossover between a molecular Bose-Einstein Condensate (BEC) and a BCS superfluid \cite{Bartenstein2004,Regal2004,Zwierlein2004,Bourdel2004}.

Recently, efforts have been made to combine reduced dimensionality with the tunability of interactions and to experimentally explore ultracold 2D Fermi gases \cite{Martiyanov2010,Feld2011,Dyke2011,Koschorreck2012,Sommer2012,Makhalov2014}. However, the phase transition to a condensed phase has not yet been observed. Here, we report on the condensation of pairs of fermions in the quasi-2D BEC-BCS crossover.

The BEC-BCS crossover smoothly links a bosonic superfluid of tightly bound diatomic molecules to a fermionic superfluid of Cooper pairs in 2D as well as 3D systems. 
However, changing the dimensionality leads to some inherent differences. In two dimensions, there is a two-body bound state for all values of the interparticle interaction. Furthermore, because of the enhanced role of fluctuations in 2D, true long-range order is forbidden for homogeneous systems at finite temperature \cite{Mermin1966,Hohenberg1967}. Still, a low temperature superfluid phase with quasi-long-range order can emerge due to the BKT mechanism \cite{Berezinskii1972,Kosterlitz1973}. 

In a 2D gas with contact interactions, the interactions can be described by the 2D scattering length $a_{2D}$.
Using the Fermi wave vector $k_F$, the dimensionless crossover parameter is given by $\ln(k_F a_{2D})$. The crossover regime is reached for $|\ln(k_F a_{2D})|\lesssim 1$.
For $\ln(k_F a_{2D})\ll -1$, the binding energy is large and the system consists of deeply bound bosonic dimers. For $\ln(k_F a_{2D})\gg 1$, the dimer binding energy tends to zero. For a thermal energy $k_B T$ significantly larger than the binding energy, the dimers are dissociated due to thermal excitations and the system becomes fermionic.

\begin{figure} [htb!]
\centering
	\includegraphics [width= 4.8cm] {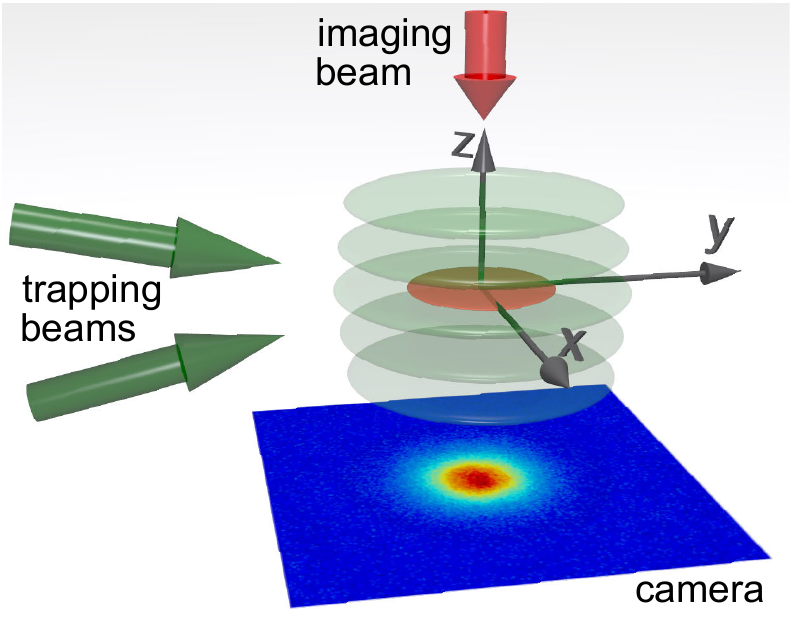}
	\caption{ \textbf{Experimental setup.} A quasi-2D gas (red disk) is created by loading a two-component ultracold Fermi gas of $^6$Li atoms into a single layer of a standing-wave trap created by two interfering laser beams ($\lambda=1064\,$nm, green arrows) that cross under a small angle ($14^{\circ}$). Using absorption imaging along the vertical direction (red arrow) we obtain the column density of the sample.}
	\label{fig:figure1}
\end{figure}

Two-dimensional gases are realized by a strongly anisotropic confinement, which leads to a freezing out of the degrees of freedom in one spatial direction. Such a quasi-2D gas captures the essential properties of a 2D system. Corrections to the 2D physics may arise from the residual influence of the third dimension.

\begin{figure*} [!htb]
\centering
	\includegraphics [width= 15.2 cm] {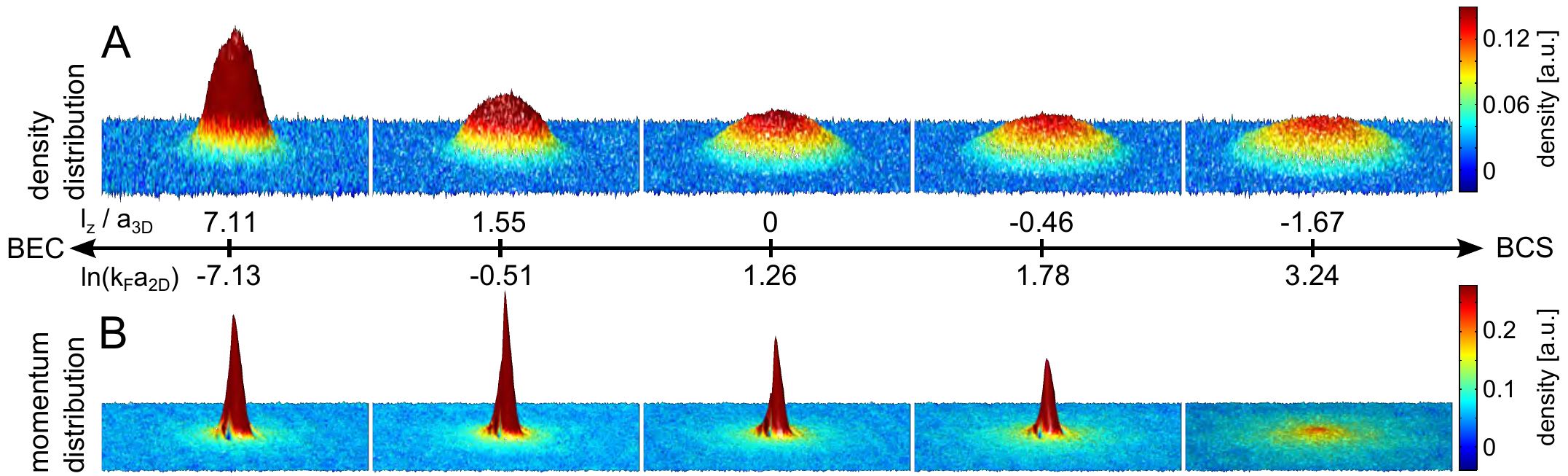}
	\caption{\textbf{Density distributions at the lowest accessible temperature for different interaction strengths.}   \textbf{(A)} In situ density distribution obtained from absorption imaging along the $z$-axis. \textbf{(B)} Pair momentum distribution obtained from the $\tau/4$ method with a pair projection ramp to $\ell_z/a_{3D}=7.11$ ($692\,$G). 
	The strong enhancement at low momenta in the momentum distribution for $\ln(k_F a_{2D})<3.24$ 
	is a clear signature of pair condensation. Each picture is the average of about $30$ individual measurements. The temperature of the samples ranges from $64\,$nK at $\ln(k_F a_{2D})=-7.13$
	to $78\,$nK at $\ln(k_F a_{2D})=3.24$.
	}
	\label{fig:figure2}
\end{figure*}

We perform our measurements using a two-component Fermi gas of $^6$Li atoms in the lowest two Zeeman sublevels, which we denote $\ket{1}$ and $\ket{2}$ \cite{Zuern2013}. The ultracold gas initially consists of $40\,000$-$50\,000\,$ 
atoms per spin state, which are bound into dimers at a temperature of approximately $50\,$nK 
and a magnetic offset field of $795\,$G ($\ell_z/a_{3D}=1.08$) \cite{SOM}. It is loaded into a hybrid trap consisting of a single layer of a standing-wave optical dipole potential and a weak magnetic potential. The combined trapping frequencies are $\omega_x=2 \pi \times 17.88(3)$\,Hz and $\omega_y= 2 \pi \times 17.82(4)$\,Hz in radial, and  $\omega_z= 2 \pi \times 5.53(3)$\,kHz in axial direction. This leads to a pancake-shaped cloud with an aspect ratio of $\frac{\omega_z}{\omega_r} \approx 310$ (see Fig.\ref{fig:figure1}) and an axial harmonic oscillator length $\ell_z = \sqrt{\hbar / m \omega_z} \approx 551\,$nm with the reduced Planck's constant $\hbar$, the atom mass $m$, and the axial trapping frequency $\omega_z$. We ensure that there is no significant population of axially excited states by measuring the axial momentum distribution of the gas \cite{SOM,Dyke2011}. 
Assuming that the internal structure of pairs, i.e. the relative wave function of the fermions inside the pairs, has only negligible effect beyond the two-body sector \footnote{For $E_B\gg \hbar \omega_z$, the system consists of tightly bound bosonic molecules whose internal structure is not resolved. For $E_B\ll \hbar \omega_z$ the internal structure of the dimers is well described within the 2D framework. Only for $E_B \approx \hbar \omega_z$, the internal structure is influenced by the third spatial dimension and can deviate from the 2D predictions. This affects the short-range behavior of the system. However, we expect the influence on the long-range behavior to be negligible (see section IV in \cite{SOM}).}, our system can be described in the 2D framework with the effective 2D scattering length $a_{2D} = \ell_z \sqrt{\pi/A} \, \exp\Bigl(-\sqrt{\frac{\pi}{2}}\frac{\ell_z}{a_{3D}}\Bigr)$ \cite{petrov2000b,Martiyanov2010,Levinsen2014,SOM}, where $A=0.905$. 

To explore the phase diagram of the quasi-2D BEC-BCS crossover, we tune the temperature by heating the sample, and the interaction strength by adiabatically ramping the magnetic offset field to values between $692\,$G ($\ell_z/a_{3D}=7.11$) and $982\,$G ($\ell_z/a_{3D}=-2.35$) \cite{SOM}.
We probe the 2D density distribution via absorption imaging along the vertical direction (see Fig. \ref{fig:figure1}). The density distributions for different interaction strengths are shown in Fig. \ref{fig:figure2}A for the coldest accessible temperatures. 
For growing $\ln(k_F a_{2D})$, the width of the sample increases while its central density decreases from approximately $2.7/\mu\text{m}^2$ at $\ln(k_F a_{2D})=-7.13$ 
to approximately $0.76/\mu\text{m}^2$ at $\ln(k_F a_{2D})=3.24$. 
This change of the density distribution illustrates the crossover from a dense condensate of bosonic molecules to a degenerate Fermi gas whose density is reduced by the Fermi pressure.
However the phase transition into a condensed phase, which manifests itself in the enhanced density of pairs with vanishing momentum, is not directly visible in the measured density distributions.

We thus conceived a method to probe the in situ pair momentum distribution of our strongly interacting system by combining a quench of interactions with a matter wave focusing technique, in which the sample expands ballistically in a weakly confining radial harmonic potential \cite{Shvarchuck2002,VanAmerongen2008,Tung2010,Murthy2014}.

Due to its large aspect ratio, our sample expands rapidly and almost exclusively in the z-direction after the release from the optical trap. Hence, its density suddenly drops and interactions between the expanding particles are quenched. Redistribution of momentum in the radial direction during the expansion is thus negligible at the weakest probed interaction strengths and does not affect the momentum distribution.
To minimize interaction effects also in the strongly interacting regime, we perform a fast ramp to the lowest accessible interaction strength on the BEC side ($B = 692\,$G, $\ell_z/a_{3D}=7.11$) on a time scale shorter than $125\,\mu$s 
just before the release. This is fast enough that the density and momentum distributions cannot adjust to the new interaction parameter \cite{SOM,Murthy2014}. 
At the same time, pairs of atoms are projected onto deeply bound molecules whose binding energy $E_B$ significantly exceeds the energy scale given by the axial confinement ($\hbar \omega_z$) and one obtains the pair momentum distribution 
\footnote{Due to this projection, information about the relative momentum of the paired atoms is lost. We therefore do not observe the Tan contact in the pair momentum distribution \cite{Tan2008}.}.
A similar technique was already used to explore the three-dimensional BEC-BCS crossover \cite{Regal2004, Zwierlein2004, Zwierlein2005}. However, these experiments could not take advantage of the interaction quench and the subsequent ballistic expansion since they were lacking the fast expansion in the z-direction.

To obtain the radial momentum distribution, we perform this ballistic expansion in a weakly confining harmonic potential with trap frequency $\omega_{\text{exp}}= 2\pi \nu_{\text{exp}}$ in the radial direction. 
In a simple picture, the harmonic potential acts as a matter wave lens and
brings the far field distribution to finite time scales. After an expansion time of $t_{\text{exp}}$=$\tau/4$, where $\tau=1/\nu_{\text{exp}}$ is the period of the harmonic potential, 
the position of each particle depends only on its initial momentum in the radial plane. Thus $n(\textbf{x}, t=\tau/4) = \tilde n (\hbar \textbf{k}/(m \omega_{\text{exp}}), t=0)$ and hence, by imaging the density profile after $t_{\text{exp}}$=$\tau/4$, we gain direct access to the initial 2D momentum distribution \cite{Shvarchuck2002,VanAmerongen2008,Murthy2014}. 
In our case, the radial trap frequency is $\omega_{\text{exp}}\simeq2\pi \times 10\,$Hz, which leads to $t_{\text{exp}} = 25\,$ms \cite{SOM}. In contrast to conventional time-of-flight expansion, where the initial spatial distribution of the sample influences the obtained momentum distribution especially at low momenta, distortions are negligible in this method.

By combining the interaction quench with the projection onto molecules and the $\tau/4$ momentum imaging, we are able to access the radial in situ pair momentum distribution $\tilde{n}(k)$ in the whole crossover regime.

\begin{figure} [!htb]
\centering
	\includegraphics [width= 6.6 cm] {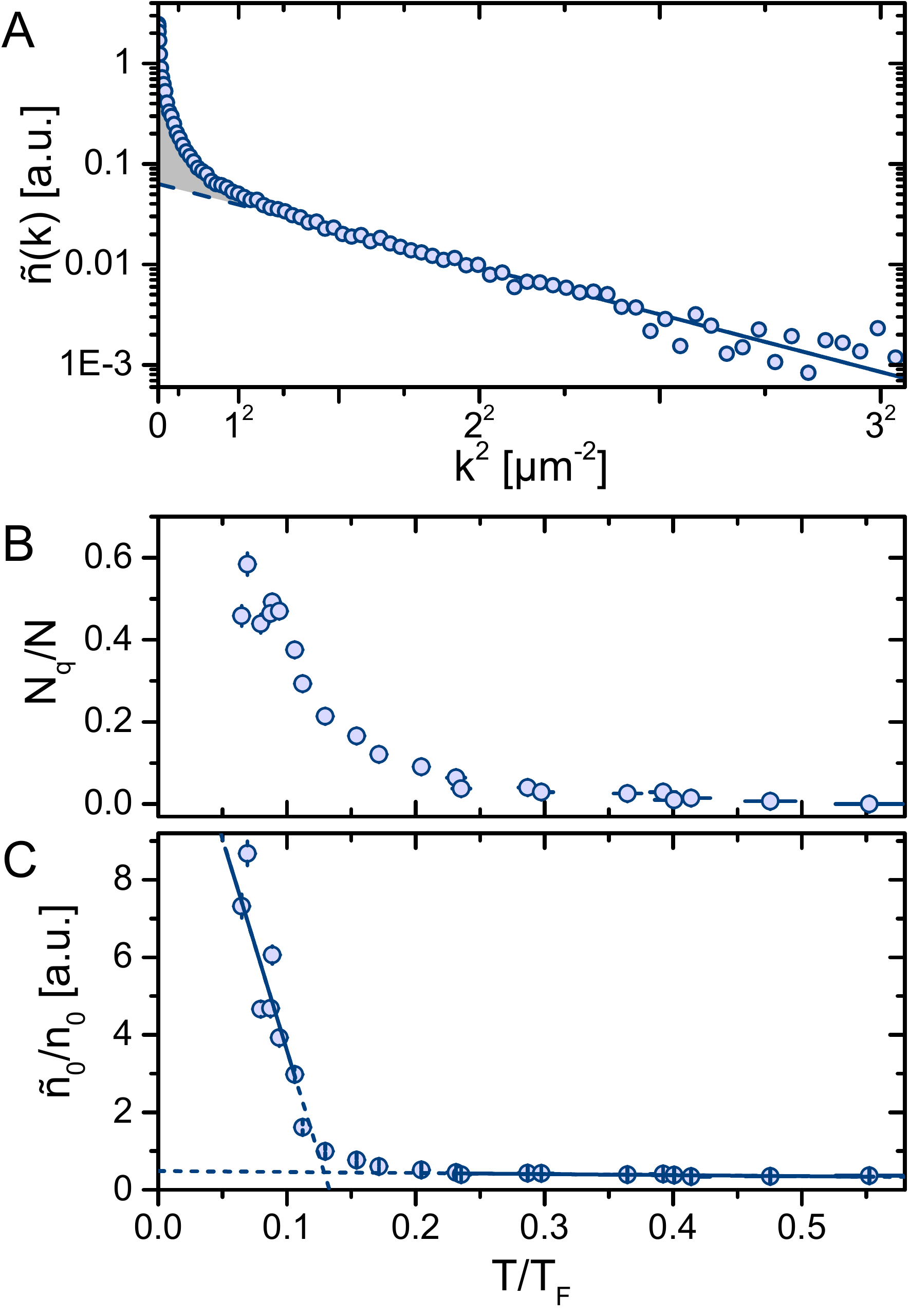}
	\caption{\textbf{Quantitative analysis of the momentum distribution at $\boldsymbol{\ell_z/a_{3D}=1.55}$.}
	\textbf{(A)} Radial momentum distribution $\tilde{n}(k)$ at the coldest accessible temperature. We logarithmically plot $\tilde{n}(k)$ as a function of $k^2$. The thermal wing thus appears as a straight line from which we extract the temperature of the sample with a Boltzmann fit (line). 
	The figure is the average of about $30$ individual measurements.
	\textbf{(B)} Nonthermal fraction $N_q/N$ as a function of $T/T_F$. $N_q$ is indicated by the gray area in panel (A).
	\textbf{(C)} Normalized peak momentum density $\tilde{n}_0/n_0$ as a function of $T/T_F$.  The intersection of linear fits to the high and low temperature regime yields the critical temperature $T_c/T_F$.
	Each data point in (B) and (C) is the average of about 30 individual measurements, the error bars indicate the standard error of the mean. Solid lines indicate the fitted data range.}
	\label{fig:figure3}
\end{figure}

Fig. \ref{fig:figure2}B shows the obtained pair momentum distributions for the coldest attainable temperature at different interaction strengths. One observes a dramatic enhancement at low momenta which manifests itself in a sharp central peak. 
This feature is strongest on the BEC side and persists above $\ln(k_F a_{2D})=0$ and the 3D Feshbach resonance, until it vanishes at $\ln(k_F a_{2D}) \approx 3.2$ 
on the BCS side.
Comparing the data at the two largest depicted values of $\ln(k_F a_{2D})$, one observes that the peak momentum density $\tilde{n}_{0}$ changes by almost an order of magnitude, whereas the in situ peak density $n_{0}$ changes by less than 10\%. As $\tilde{n}_{0}$ is a measure for the long-range coherence of the system \cite{Plisson2011}, the observed abrupt change indicates the phase transition to the condensed phase.

For a more quantitative analysis of our data, we azimuthally average the pair momentum distribution. 
Fig. \ref{fig:figure3}A shows the obtained radial distribution for the coldest accessible temperature measured at $782\,$G, which corresponds to $\ell_z/a_{3D}=1.55$ ($\ln(k_F a_{2D})\approx - 0.51$). 
We extract the temperature $T$ of the sample from the high momentum tail of the radial distribution which is well described by a Gaussian.
Note that before the ramp of the interaction strength, the thermal part of the gas consists of molecules for $\ell_z/a_{3D}>0.55$, free atoms for $\ell_z/a_{3D}<-0.46$, and a mixture of atoms and molecules between these two interaction strengths \cite{SOM}.
For each investigated interaction strength and temperature, we determine the Fermi wave vector and Fermi temperature from the in situ peak density according to 
$k_F^2 = 2 m k_B T_F / \hbar^2 = 4 \pi n_{0}$.
Here, $m$ refers to the mass of a $^6$Li atom and $k_B$ is Boltzmann's constant. This definition employs the local density approximation at the trap center and allows us to compare the obtained data to predictions for the homogeneous system. Note that $n_{0}=n_{0,{\vert1\rangle}}=n_{0,{\vert2\rangle}}$, where $n_{0,{\vert i\rangle}}$ is the peak density of atoms in state $\vert i \rangle$.

At low momenta, a fraction of the momentum density lies above the Gaussian fit (gray area in Fig. \ref{fig:figure3}A). We define this quantity as the nonthermal fraction $N_q / N$ \footnote{We identify the non-Gaussian fraction $N_q/N$ with the fraction of the cloud which has non-Gaussian fluctuations. In the literature \cite{Prokofev2001,Prokofev2002}, this is referred to as the quasicondensate. Below $T_c$, the quasicondensate density becomes identical to the superfluid density in mean-field theory.} and investigate its behavior as a function of the degeneracy temperature $T/T_F$ (see Fig. \ref{fig:figure3}B).
While the non-Gaussian fraction vanishes for $T / T_F \gtrsim 0.5$, it slowly grows for decreasing temperatures. For $T / T_F \lesssim 0.2$, the slope increases until we reach $N_q / N \approx 0.6$ for the coldest samples. This is in agreement with theoretical predictions \cite{Prokofev2001,Prokofev2002,Bisset2009} and previous experimental results \cite{Plisson2011,Tung2010,Clade2009,Hung2011a}, which find a presuperfluid increase of low-momentum states for temperatures above the superfluid transition temperature $T_c$. This behavior inhibits a precise determination of the transition temperature $T_c$ from $N_q / N$. 
To obtain an estimate for the critical temperature, we instead plot the normalized peak momentum density $\tilde{n}_{0}/n_0$ as a function of temperature as shown in Fig. \ref{fig:figure3}C.
This quantity is a measure for the fraction of the sample which exhibits long-range phase coherence \cite{Plisson2011}. The innermost pixel of the momentum distribution corresponds to a coherence length well above $100\,\mu$m 
which is almost two orders of magnitude larger than the thermal wavelength of the coldest samples. The normalized peak momentum density
shows a sudden change of slope which we assume to occur at the phase transition. We estimate $T_c/T_F$ by the intersection of linear fits to the regimes above and below the phase transition. 
For the example shown in  Fig. \ref{fig:figure3}, this results in a critical temperature of $T_c/T_F=0.129 \,(35)$, 
where the statistical uncertainty is obtained from the standard errors of the two linear fits. The critical phase space density is $\rho_c = n_{0,c} \lambda_{dB,c}^2 = 3.9 \,(6)$, 
where $\lambda_{dB,c}$ and $n_{0,c}$ are the thermal de Broglie wavelength and the peak in situ density at the critical temperature, respectively. 

\begin{figure} [!htb]
\centering
	\includegraphics [width= 9. cm] {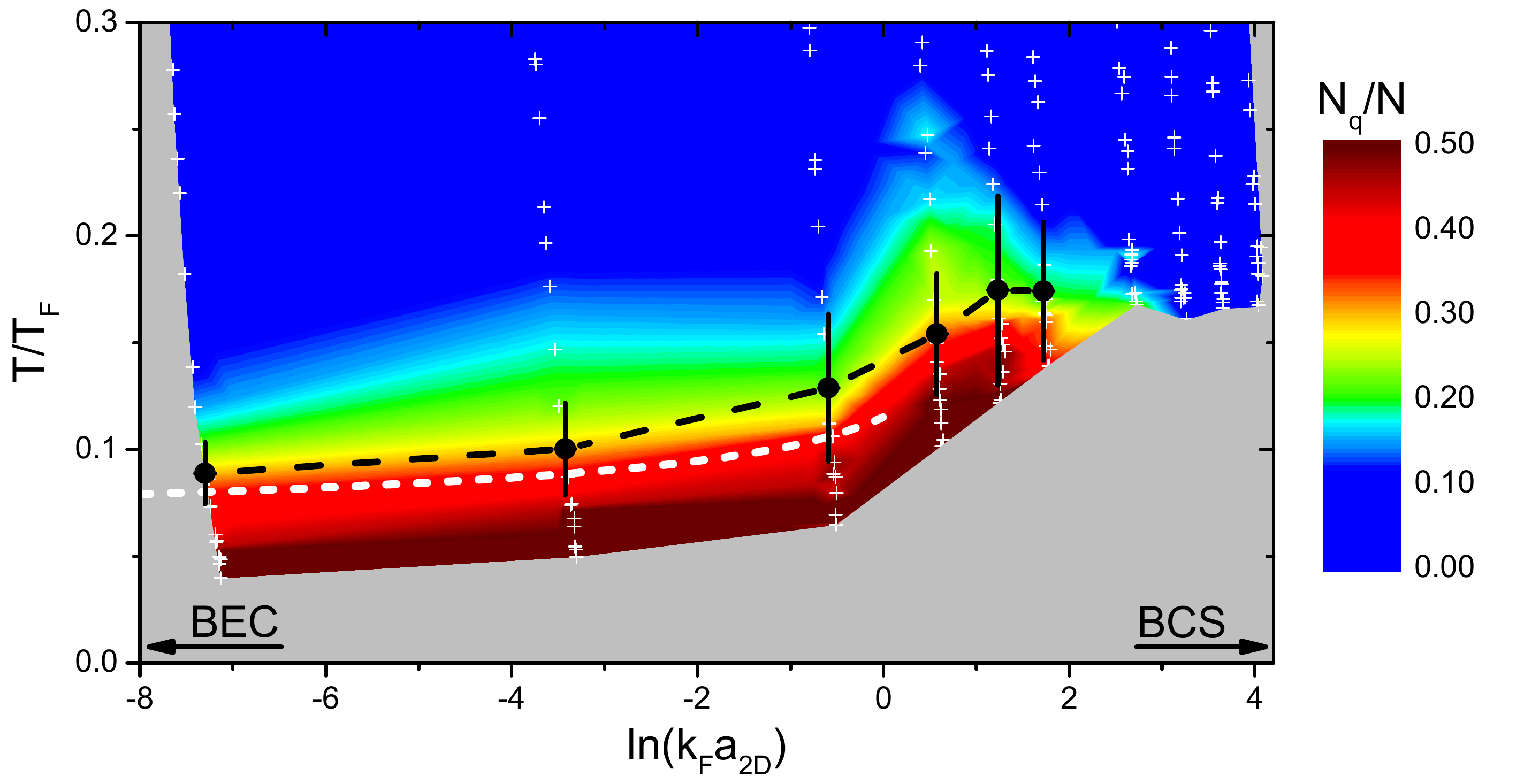}
	\caption{\textbf{Phase diagram of the strongly interacting 2D Fermi gas.} The experimentally determined critical temperature $T_c/T_F$ is shown as black data points and the error bars indicate the statistical errors. Systematic uncertainties are discussed in detail in \cite{SOM}. The color scale indicates the nonthermal fraction $N_q / N$ and is linearly interpolated between the measured data points (white crosses). Each data point is the average of about $30$ measurements. The dashed white line is the theoretical prediction for the BKT transition temperature given in \cite{Petrov2003}. }
	\label{fig:figure4}
\end{figure}

By repeating this analysis for all investigated interaction strengths, we obtain the transition temperature as a function of the interaction parameter $\ln(k_F a_{2D})$. The resulting values are shown as black dots in Fig.\ref{fig:figure4} together with the corresponding non-Gaussian fraction $N_q/N$, which is displayed as a color scale. Comparing the data for $T_c/T_F$ and $N_q/N$, one finds that the phase transition occurs at a significant non-Gaussian fraction of $N_q/N \approx 0.3$ for all measured interaction strengths.

On the BEC side of the phase diagram, one observes a slow increase of the measured critical temperature towards the crossover region.
Within their statistical uncertainties, the measured values of $T_c/T_F$ are in good agreement with an effective description in terms of 2D bosons \cite{Petrov2003}. 
This theoretical prediction describes a BKT transition into a superfluid phase 
with algebraically decaying phase coherence.
Interestingly, the bosonic theory  provides a reasonable description of the data up to $\ln(k_Fa_{2D})=0$, where the 2D scattering amplitude diverges. 
This indicates that the fermionic nature of the constituents of the bosonic dimers has only a small effect on the many-body physics of the system up to this point. The crossover to a fermionic description should thus occur at positive values of $\ln(k_Fa_{2D})$. This is in line with recent theoretical predictions \cite{Bertaina2011,Ngampruetikorn2013}.

Far on the BCS side, fermionic theories predict an exponential decrease of $T_c/T_F$ \cite{SadeMelo1993,Bauer2014}. Although we can only give an upper limit for the critical temperature $T_c/T_F \leq 0.16$ 
for $\ln(k_F a_{2D})\geq 2$, the observed non-Gaussian fraction is consistent with a decrease towards the BCS limit. 
However, $T_c/T_F$ is systematically above the theoretical predictions for $\ln(k_F a_{2D})>0$ 
\cite{Petrov2003,Bauer2014,Matsumoto2014}. Part of this deviation might be due to 
the residual influence of the third dimension. In our system, residual axial excitations grow with increasing $\ln(k_F a_{2D})$ \cite{SOM}. Recently, it was predicted that they would lead to an increased critical temperature \cite{Fischer2014}. 
Additionally, the three-dimensional internal structure of atom pairs might lead to corrections in the regime where $E_B \approx \hbar \omega_z$, which go beyond the two-body sector. Whether this effect has any influence on the measured phase diagram still needs further experimental and theoretical consideration. Initial steps in this direction have been taken \cite{Dyke2014}.

Our work constitutes a basis for future theoretical and experimental studies of quantum gases in the quasi-2D BEC-BCS crossover. The measured critical temperature suggests the validity of BKT theory on the bosonic side. Superfluidity and the algebraic decay of correlations below the transition remain to be validated. Indeed, our ability to extract the in situ momentum distribution with negligible distortion offers direct access to the coherence properties of the system. A first analysis of the trap averaged first order correlation function, which we obtain by Fourier transforming the pair momentum distribution, suggests algebraically decaying phase correlations below the critical temperature. However, due to the inhomogeneity of our system, a careful analysis is required to unambiguously confirm the BKT nature of the observed transition. Additionally, the equation of state can be extracted from the density distribution in the trap. 
Finally, the exploration of the dimensional crossover to 3D, in which an increased $T_c/T_F$ is predicted \cite{Fischer2014}, offers new opportunities to understand mechanisms which lead to high critical temperatures.


\begin{acknowledgments}
The authors would like to thank T.~Enss, J.~Levinsen, P.~Massignan, L.~Mathey, M.M.~Parish, and J.M.~Pawlowski for valuable discussions, and J.H.~Becher, J.E.~Bohn, and S.~Pres for contributions to the construction of the 2D setup.
They gratefully acknowledge support from ERC Starting Grant No. $279697$, ERC Advanced Grant No. $290623$, the Helmholtz Alliance HA$216/$EMMI, and the Heidelberg Center for Quantum Dynamics. M. G. R. and I. B. acknowledge support by the Landesgraduiertenf\"orderung Baden-W\"urttemberg.
\end{acknowledgments}

\cleardoublepage

\section*{SUPPLEMENTAL MATERIAL}
\setcounter{figure}{0}
\renewcommand{\figurename}{Fig.\,S}

\section{Two-dimensional interactions and theoretical predictions}

In our experiment we create an effectively 2D system by means of a tight axial confinement along the $z$-direction. The system is then characterized by the 3D scattering length $a_{3D}$ and, as an additional length scale, the oscillator length $\ell_z=\sqrt{\hbar/m\omega_z}$. Both parameters uniquely determine the effective 2D scattering length $a_{2D}$ of the system. In our measurements $\ell_z$ is constant and we tune $a_{2D}$ by changing $a_{3D}$ via a magnetic Feshbach resonance. 
Unlike in 3D, there is a two-body bound state for all interaction strengths in two dimensions. In the BCS limit ($a_{2D} \gg \ell_z$), it is weakly bound and the size of the dimers, which is on the order of $a_{2D}$, is much larger than $\ell_z$. Thus, the dimers 
have a 2D character. On the contrary, in the BEC limit the dimers are much smaller than $\ell_z$. They are thus not influenced by the axial confinement, and internally have a 3D character. However, the behavior of the dimers can be described in the 2D framework regardless of their internal structure. In particular, the scattering processes in this system 
can be mapped onto those of purely two-dimensional interactions. 
We review here the concepts of 2D scattering length and confinement induced two-body bound state.

\textbf{2D scattering length.} The 2D scattering length $a_{2D}$ is defined from low-energy scattering of particles \cite{Bloch2008,Makhalov2014,Levinsen2014}. For this purpose we consider the scattering amplitude of two atoms with relative momentum $\hbar k$ and energy $E_k=\hbar \omega_z/2+\hbar^2k^2/m$ colliding in the continuum above the ground state of the harmonic oscillator in $z$-direction. For sufficiently low $k$ we have $|E_k-\hbar\omega_z/2|\ll \hbar\omega_z$ and higher excitations in the $z$-direction do not affect the collision. This regime is experimentally relevant here for all magnetic fields since $\mu,k_B T\lesssim \hbar\omega_z$. The scattering amplitude for atoms then reads
\begin{align}
 \label{a2d1} f(k) = \frac{4\pi}{\sqrt{2\pi} \ell_z/a_{3D}-\ln(\pi k^2\ell_z^2/A)+\mathrm{i} \pi}
\end{align}
with $A=0.905$ \cite{Petrov2001,Bloch2008}. This formula is valid for all values of $\ell_z/a_{3D}$ as long as $k$ is sufficiently low \cite{Bloch2008,Levinsen2014}. In particular, from the definition of the 2D scattering length $a_{2D}$ according to $f(k) \to 4\pi/[-\ln(k^2a_{2D}^2)+\mathrm{i} \pi]$ for $k\to 0$ we deduce
\begin{align}
 \label{a2d2} a_{2D} = \ell_z \sqrt{\frac{\pi}{A}} \exp\Bigl(-\sqrt{\frac{\pi}{2}}\frac{\ell_z}{a_{3D}}\Bigr)
\end{align}
for all magnetic field values. 
This definition includes the underlying three-dimensionality of the scattering events up to the two-body sector.

In the weak confinement limit ($\ell_z \gg a_{3D}$) realized far on the BEC side, we obtain
\begin{align}
 \label{a2d3} f(k) \simeq \tilde{g} := \sqrt{8\pi} \ \frac{a_{3D}}{\ell_z},
\end{align}
which constitutes an energy-independent effective coupling constant. This is the expected low-energy limiting behavior for a short ranged potential whose  range $r_e$ is much smaller than $\ell_z$ \cite{Bloch2008}. Expression (\ref{a2d3}) can be used to compare the molecules on the BEC side to two-dimensional bosons with coupling constant $\tilde{g}_b$ by replacing $a_{3D} \to 0.6 a_{3D}$ \cite{Petrov2005} and $m\to 2m$. We then obtain $\tilde{g}_b=0.60$ for $692$G, demonstrating that the fermionic system realizes strongly coupled two-dimensional bosons in this limit. Eq. (\ref{a2d2}) for the 2D scattering length has been employed for the whole crossover in Ref. \cite{Makhalov2014}, where the limit (\ref{a2d3}) provides for the correct bosonic weak-coupling limit.

\textbf{Confinement induced bound state.} The confinement along the $z$-direction induces a two-body bound state with binding energy $\tilde{E}_B$ for all values of the 3D scattering length $a_{3D}$. The corresponding binding energy $\tilde{E}_B$ is found from \cite{Petrov2001,Bloch2008}
\begin{align}
 \frac{\ell_z}{a_{3D}} =\int_0^\infty \frac{\mbox{d}u}{\sqrt{4\pi u^3}} \Bigl(1-\frac{e^{-\tilde{E}_Bu/\hbar\omega_z}}{\sqrt{(1-e^{-2u})/2u}}\Bigr).
 \label{EqTrans}
\end{align}
From the binding energy an associated length scale $\tilde{a}_{2D}$ can be defined according to $\tilde{E}_B=\hbar^2/m\tilde{a}_{2D}^2$. This length scale can be used to quantify the interactions of the system. However, only far on the BCS side of the crossover, where ${a}_{2D}\gg\ell_z$ and the internal structure of the dimer is  two-dimensional, this length scale coincides with the 2D scattering length. 

This limit is reached for magnetic fields $\gtrsim 852$G. Far on the BEC side, in contrast, $\tilde{E}_B$ approaches the binding energy of the 3D system since the size of dimer is much smaller than $\ell_z$ and thus the effect of the strong confinement on the dimer vanishes. The difference between $a_{2D}$ and $\tilde{a}_{2D}$ reaches almost three orders of magnitude at $692$G. The interaction strengths at the critical temperature using both definitions are listed in Table I.

Far on the BEC side ($\ln(k_Fa_{2D})\ll-1$), the critical temperature can be computed from the theory of weakly coupled two-dimensional bosons. Based on the corresponding Monte Carlo calculations presented in \cite{Prokofev2001,Prokofev2002} the BKT transition temperature is found \cite{Petrov2003} to be
\begin{align}
 \frac{T_c}{T_F} = \frac{1}{2}\Bigl[\ln\Bigl(\frac{C}{4\pi}\ln\Bigl(\frac{4\pi}{k_F^2a_{2D}^2}\Bigr)\Bigr)\Bigr]^{-1}
\end{align}
with $C=380(3)$.

\section{Preparation of the sample}

We start our experimental sequence by transferring a 3D Fermi gas of $^6$Li atoms in states $\ket{1}$ and $\ket{2}$ \cite{Zuern2013} from a magneto-optical trap into an optical dipole trap (ODT). This surfboard-shaped trap has an aspect ratio of  $\omega_x$:$\omega_y$:$\omega_z = 1$:$8$:$44$  and is far red detuned ($\lambda = 1064\,$nm) from the optical transition. 
The gas is then evaporatively cooled into degeneracy at a magnetic offset field of $795\,$G on the BEC side of the broad Feshbach resonance at $832.2\,$G \cite{Zuern2013}. We therefore obtain a 3D molecular Bose-Einstein condensate (mBEC) consisting of about $10^5$ molecules with negligible thermal fraction. This sample is finally transferred into a standing-wave optical dipole trap (SWT) as illustrated in Fig. 1A in the main text.

The SWT is created by two elliptical focused $1064\,$nm Gaussian beams, which intersect under an angle of $\simeq 14^\circ$. This leads to a standing wave interference pattern where the maxima have a distance of $\simeq4.4\,\mu$m.  The ellipticity of the beams is chosen such that the interference maxima have a circular symmetry in the xy-plane. At the position of the SWT, the magnetic offset field has a saddle point. It leads to an additional weak magnetic confinement (anti-confinement) in radial (axial) direction. The measured trap frequency of the magnetic confinement in radial direction is $\omega_{mag}(B) \approx 2\pi \times 0.39 \text{Hz} \, \sqrt{B [\text{G}]}$.
At a magnetic offset field of $795\,$G, the combined trapping frequencies for the central layers of the SWT are given in the main text and lead to an aspect ratio of $\omega_x:\omega_y:\omega_z = 1:0.997:309$. 

In order to align the relative position of the atoms in the ODT with one layer of the SWT, we apply a magnetic field gradient 
in z-direction, which can shift the atoms up or down in the ODT. 
To optimize the fraction of atoms transferred into this single layer, we furthermore decrease the vertical size of the atoms in the ODT by modulating the position of the ODT in the transverse x-direction on a time scale much faster than all trapping frequencies. This creates a time averaged potential where the width of the trap in x-direction is increased by a factor of approximately 5. 
In order to further reduce the extension in z-direction of the sample, we ramp to a magnetic offset field of $730\,$G over $600\,$ms. This reduces the repulsive interaction and thus the size of the sample. After the transfer into the SWT, we ramp back to $795\,$G, where we further evaporatively cool the sample by simultaneously applying a magnetic field gradient and lowering the trap depth. This also allows us to control the total number of particles.

In order to access higher temperatures in a controlled fashion, we can then apply a heating procedure. For the lowest three temperatures, we hold the sample in the SWT at $795\,$G for a variable time ($0 \dots 1\,$s) during which it is heated by technical noise. For higher temperatures, we parametrically heat the sample by modulating the depth of the SWT with variable amplitude.  After letting the sample equilibrate for $300\,$ms, we ramp the magnetic offset field to the value we want to investigate, where we wait for an additional $20\,$ms before probing the system. All magnetic field ramps are performed with ramp speeds $\lesssim 1.9\,$G/ms. 
To ensure adiabaticity of the magnetic field ramps, we compare the temperature of a sample held at an offset field of $732\,$G to the temperature of a sample which was ramped across the Feshbach resonance to $900\,$G and back in the same time. We find that for this ramp speed the two temperatures agree within their uncertainties. All magnetic field ramps are thus adiabatic, and we probe the crossover in an isentropic way.

\section{Distribution of particles in the standing-wave trap}
\label{subsec:tomography}

\begin{figure} [!htb]
\centering
	\includegraphics [width= 7 cm] {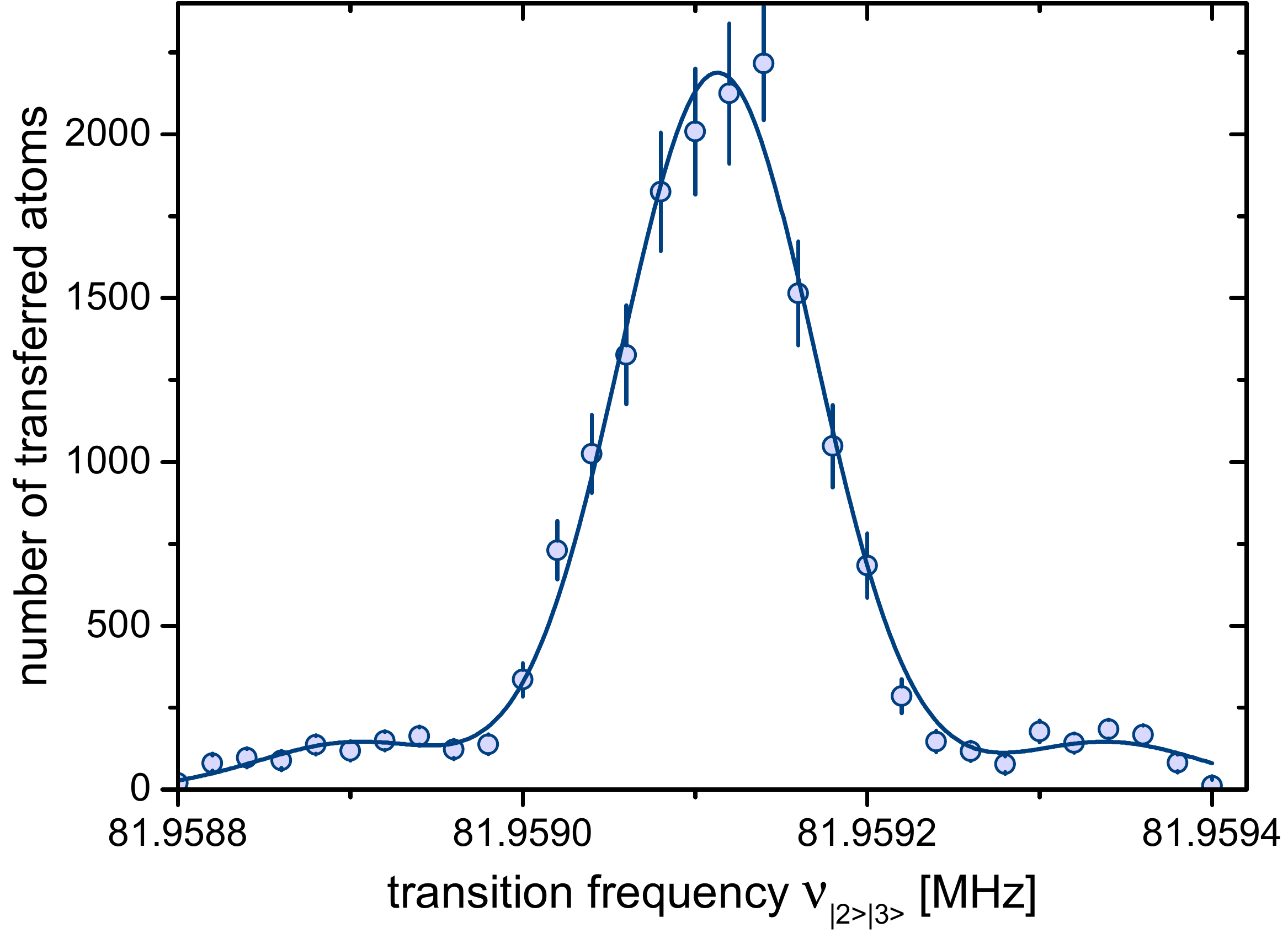}
	\caption{\textbf{Tomographic measurement of the particle distribution in the standing-wave trap.} Data points represent the number of particles transferred to state $\ket{3}$ as a function of the transition frequency $\nu_{\ket{2}\ket{3}}$.
	The central maximum at $81.9591\,$MHz corresponds to atoms in the central layer, the neighboring layers are only slightly populated. The sum of three Gaussian profiles (solid line) is fitted to the data and yields a population of the central layer with approximately $89$\% of the particles.}
	\label{fig:figures1}
\end{figure}

In order to probe the distribution in the layers of the SWT, we use a radio-frequency tomography technique. We apply a magnetic field gradient along the z-axis to make the 
transition frequency $\nu_{\ket{2}\ket{3}}$  between states $\ket{2}$ and $\ket{3}$ spatially dependent on z. The dependence of $\nu_{\ket{2}\ket{3}}$ on the magnetic field is given by $d\nu_{\ket{2}\ket{3}}/dB\simeq 6.3 \,$kHz/G. 
We can thus visualize the density distribution by counting the number of transferred atoms as a function of the transition frequency. To minimize the line width of the transition, we need to exclude interaction effects and three-body losses. Hence, we first remove the particles in state $\ket{1}$ by applying a resonant laser pulse for about $10\,\mu$s. To minimize heating and losses, this is done at a magnetic offset field of $1000\,$G, where the atoms are not bound into molecules at our temperatures and interactions are comparatively weak. Although we still observe significant heating, the thermal energy is small compared to the trap depth and it is therefore not expected that particles get transferred between the individual layers.
After a ramp back to $795\,$G, we apply a magnetic field gradient of approximately $70\,$G/cm along the z-axis. We thus achieve a difference of approximately $200\,$Hz in transition frequency between atoms in neighboring layers. We then drive the $\ket{2}$-$\ket{3}$ transition and record the number of transferred particles using state-selective absorption imaging along the y-axis as a function of frequency (see Fig. S\ref{fig:figures1}).

The large central maximum at $81.9591\,$MHz in Fig. S\ref{fig:figures1} corresponds to atoms in the central layer. Note that only a fraction of atoms is transferred to state $\ket{3}$, and thus the displayed atom number is considerably lower than the total atom number in the trap. The neighboring layers lie at roughly $81.9589\,$MHz and $81.9593\,$MHz which was confirmed in previous measurements where several layers were filled. By repeating the tomographic measurement, we can assure that the position of the layers is stable within a range of $\pi/8$ over time scales of more than a week.

We fit the distribution shown in Fig. S\ref{fig:figures1} with three Gaussian profiles of the same width and thus estimate the fraction of atoms in the non-central peaks to be $11$\%. Note that this value is a conservative upper bound and overestimates the number of atoms in the non-central layer for two reasons: the magnetic field gradient applied during the measurement tilts the trap and thus removes a large percentage (approximately $25$\%) of all atoms before the transition is driven. Since the central layer is filled with more atoms and to higher energies than the surrounding layers, a greater fraction of atoms will be lost from the central layer. In addition, the atom numbers detected in the non-central peaks are at the detection limit, and are thus influenced by phenomena such as dispersive non-resonant interactions between the imaging light and atoms in state $\ket{2}$.

The phase space density of atoms in the non-central layers is low, and we therefore expect them to follow a thermal distribution. Hence, their influence on the measured condensate fraction, peak condensate density and temperature is negligible. However, the in situ density distribution may be influenced. This is discussed in section \ref{subsec:systematic_errors}.

\section{Influence of the finite aspect ratio}

In contrast to theory, where the dimensionality of a system can easily be set, experimental realizations of low dimensionality will always remain an approximation. For instance, a two-dimensional system can be realized by strongly confining particles in one of the three spatial dimensions. However, there will always be a residual influence of the third dimension. Its magnitude can be determined by comparing the relevant energy scales of the system (the temperature $T$, the chemical potential $\mu$ and dimer binding energy $E_B$) to the axial oscillator energy $\hbar \omega_z$. 
For $T, \mu \gtrsim \hbar \omega_z$, particles populate the axially excited trap levels. We ensure the absence of a significant population of these excited levels by performing a measurement which is explained below. This includes the center-of-mass motion of atom pairs. However, depending on the dimer binding energy $E_B$ the internal structure of atom pairs can be three-dimensional. For $E_B\ll \hbar \omega_z$, the internal structure of the pairs is 2D. For $E_B\gg \hbar \omega_z$, the pairs are deeply bound and their internal 3D structure is not resolved. Hence, only for $E_B \approx \hbar \omega_z$ which in our system occurs at an interaction strength $\ln(k_Fa_{2D})\approx 0.5$, the internal structure of the pairs might affect the behavior of the system. 

Estimating this effect is complicated: a theoretical treatment beyond the two-body sector is extremely difficult due to strong interactions, and experimental studies would require even larger trap aspect ratios or smaller atom numbers, both of which are currently unfeasible.

We estimate the population of axially excited states due to finite $T$ and $\mu$ by investigating the system's momentum distribution in the axial direction. This measurement is performed similar to the technique described in \cite{Dyke2011} and it relies on the same principles as the previously used band-mapping technique \cite{froehlich2011}. 

\begin{figure} [!htb]
\centering
	\includegraphics [width= 7 cm] {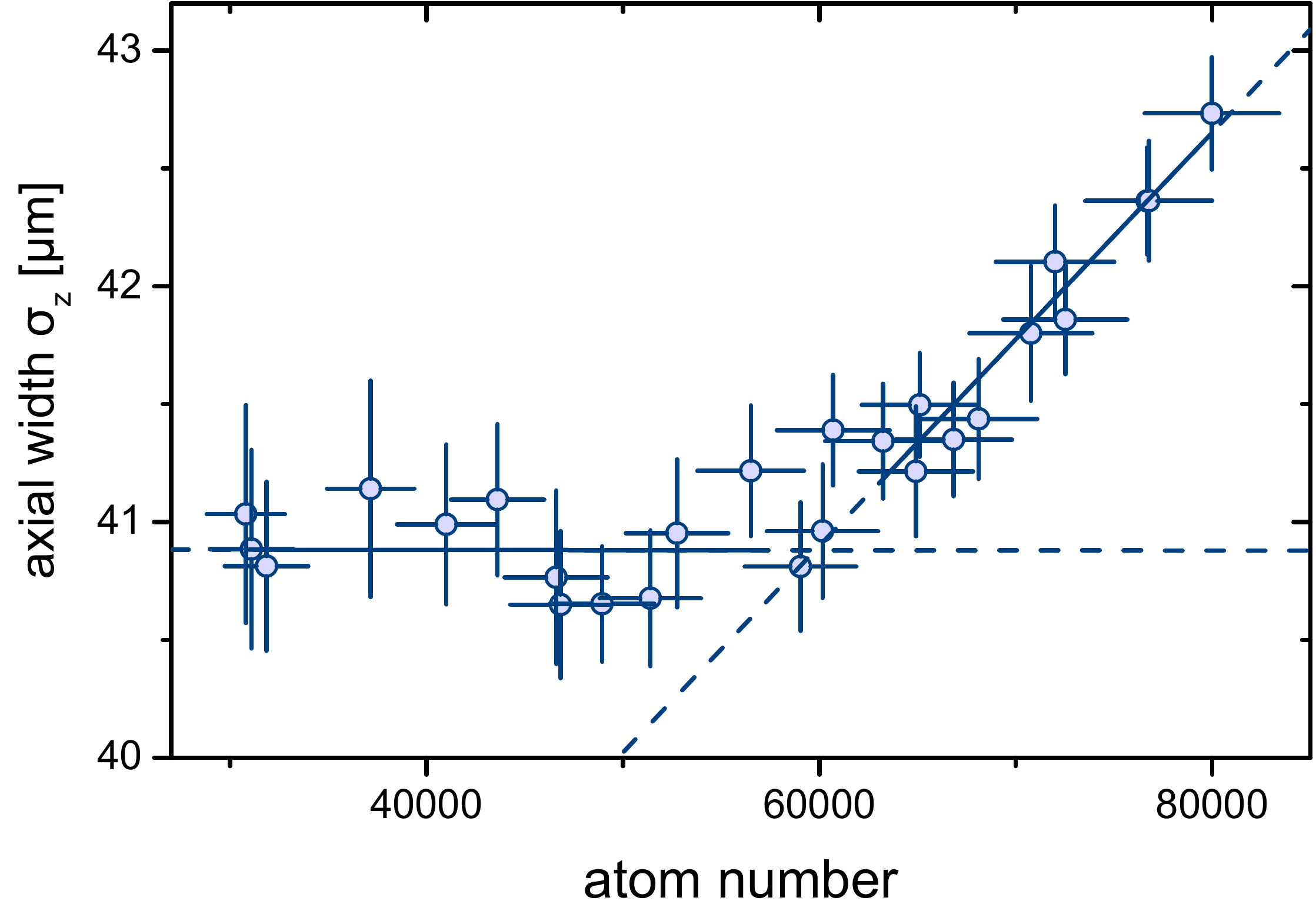}
	\caption{\textbf{Measurement of axially excited population.} The axial width $\sigma_z$ is obtained from a Gaussian fit to the density distribution after $3\,$ms time-of-flight. For atom numbers up to approximately $60\,000$, 
	only the axial ground state of the trap is occupied and $\sigma_z$ is constant. For higher atom numbers, axially excited trap levels become populated and $\sigma_z$ increases. Lines are linear fits to the data, the fit range is indicated by the solid part of each line.}
	\label{fig:figures2}
\end{figure}

We release the sample from the SWT, let it expand for a time-of-flight of $3\,$ms, and measure its vertical extension. In the non-interacting limit, atoms in the axial ground state of the trap have a Gaussian wave function in the axial direction. Their axial expansion can then be described by the dispersion of a Gaussian wave packet, which is independent of the number of atoms in the axial ground state. Fig. S\ref{fig:figures2} shows the axial width $\sigma_z$, which is determined from a Gaussian fit to the density distribution, 
as a function of the number of prepared atoms per spin state $N$ in the weakly interacting Fermi regime at $1400\,$G and at the coldest attainable temperature. One observes that the axial width is independent of the number of atoms up to approximately $N_{2D}=60\,000$ 
atoms per spin state. For $N>N_{2D}$ the axial width starts to increase with growing $N$. This change in behavior indicates population of the axially excited states, where the atoms have additional momentum in axial direction. The obtained critical atom number $N_{2D}$ is in agreement with the expectations for a trap with the given aspect ratio and anharmonicities. By keeping the atom number below $N_{2D}$, we can thus ensure that for the investigated temperature only a negligible fraction of atoms populates the axial excited state. 

This measurement is performed in the fermionic regime, where due to the Pauli principle multiple occupation of trap levels is suppressed.
All other measurements presented here are performed at lower magnetic fields, closer to the bosonic limit ($\ln(k_F a_{2D}) \rightarrow -\infty$). 
The measurement performed at $1400\,$G ($\ln(k_F a_{2D})\gtrsim6$) thus represents an upper bound on the fraction of particles in the axially excited states as for lower fields the atoms tend to form molecules which occupy lower energy states. 

To ensure the absence of a significant amount of axial excitations also for lower magnetic fields and higher temperatures, we make use of the relation between the radial size of a harmonically trapped gas and the energy of the highest occupied oscillator level.
We compare the radial size of each sample to that of the measurement at $1400\,$G, where we have excluded significant population of axially excited states. For this comparison, we estimate the radial Fermi radius $r_F$ by the radius where the particle density reaches the noise floor for a particle number $N_{2D}$ at a magnetic field of $1400\,$G. Then we integrate the number of particles outside this radius. In this way, we obtain $N_{ex,r}$, an estimate for the number of particles whose energy is larger than $\hbar \omega_z$. Assuming that all degrees of freedom are equally occupied, the number of particles in axially excited states is given by $N_{ex,z} \simeq N_{ex,r}/2$. We find that $N_{ex,z} \lesssim 1.5\%$ for all interactions strengths below $\ln(k_F a_{2D}) = 3$ and temperatures below $T/T_F \simeq 0.3$. Furthermore, for all $T/T_F \lesssim 0.2$, the fraction of axially excited particles is $N_{ex,z} < 1\%$.

Note that this estimate for the axially excited fraction is conservative, as it assumes constant trap frequencies. It does not take into account that the radial trap frequency decreases by up to about $16$\% when the magnetic field is decreased to $692\,$G.  
This leads to an increase in the corresponding Fermi radius by up to approximately $8$\%, and to a corresponding overestimation of $N_{ex,z}$.

We have thus measured the phase diagram of a quasi-2D system with small but finite influence of the third dimension. This influence has to be considered when comparing the experimental data to true 2D predictions. 
Recent theoretical work shows that these effects influence the system and can lead to a higher critical temperature \cite{Fischer2014}.

\section{Obtaining the pair momentum distribution}
\label{subsec:momentumdistribution}

In order to probe the momentum distribution of the sample, we use the combination of an interaction quench and a matter wave focusing technique described in detail in the main text and in \cite{Murthy2014}. We turn off the optical SWT, and let the sample expand ballistically in the weak magnetic potential, which is confining in the radial direction. Due to the harmonic shape of this potential, the in situ momentum distribution of the 2D sample is mapped to a spatial distribution after an expansion time of  $t_{\text{exp}}=T/4=\pi/(2\omega_{\text{exp}})$. In our case, $\omega_{\text{exp}} = \omega_{\text{mag}} \approx 2\pi \times 10\,$Hz, which leads to $t_{\text{exp}} = 25\,$ms. 

To obtain the actual in situ momentum distribution, it is fundamentally important that interactions are negligible while the gas is expanding, since they would result in a redistribution of momentum. Due to its large aspect ratio, our sample expands rapidly in z-direction. Thus, its density drops rapidly and interactions between the expanding particles are quenched. However, for large interactions strengths there is still residual scattering, which can affect the obtained momentum distribution.
We thus minimize the interactions by quickly ramping to the lowest accessible interaction strength on the BEC side ($692\,$G) on a timescale shorter than $125\,\mu$s just before releasing the sample from the SWT. This procedure leads to a negligible scattering rate during the expansion \cite{Murthy2014} and projects correlated pairs of atoms onto tightly bound molecules. The measured momentum distribution thus does not contain the relative momentum of the atoms in a pair, but only the center-of-mass momentum of the pair. Thus, fermionic Cooper pairs and bosonic molecules yield the same signature of enhanced low-momentum density in the pair momentum distribution. As a consequence, information about the Tan contact \cite{Tan2008} cannot be obtained from the pair momentum distribution.

We validate that this method does not alter the temperature of the system by comparing the momentum distributions obtained with and without the interaction quench both below ($732\,$G) and above ($872\,$G) the resonance. In both cases the observed temperature are consistent within the experimental uncertainties.

Furthermore, we confirm that the measured non-Gaussian fraction $N_q/N$ remains unchanged for low magnetic fields where pairs are deeply bound. This is achieved by comparing data with and without the magnetic field ramp. For high magnetic fields however, we cannot directly access $N_q/N$ without projecting correlated pairs into molecules. We thus need to make sure that we probe the properties of the interacting system at the original magnetic field, i.e. that the sample does not adapt to the interaction strengths at lower magnetic fields during and after the ramp.
To estimate this effect, we prepare a sample at $900\,$G at a high temperature where we expect $N_q/N = 0$ and perform the rapid ramp without releasing the sample from the trap. We find that it takes more than $11\,$ms for the momentum distribution to adapt to the interaction strength at the new magnetic field value of $692\,$G and develop a non-Gaussian fraction.
This is two orders of magnitude larger than the timescale of the rapid ramp ($< 125 \,\mu$s). Hence, the influence of the rapid ramp technique on the measured quantities can be neglected.

\section{Absorption imaging parameters and calibrations}
\label{subsec:imaging}
We use absorption imaging along the z-axis (see Fig.1, main text) to determine the integrated column density $n_{2D}\left(x,y\right)$. To obtain a reasonable signal-to-noise ratio we set the imaging intensity to $I \simeq I_{sat}$. Thus, for zero detuning one obtains  \cite{Reinaudi2007,Yefsah2011} 
\begin{align}
n_{2D}\left(x,y\right) \sigma_0^* &\,=\, - \ln{\frac{I_t\left(x,y\right)}{I_0\left(x,y\right)}}+\frac{I_0\left(x,y\right)-I_t\left(x,y\right)}{I_{sat}^*} \label{eq:imaging}\\ 
&\,=\, OD\left(x,y\right)+\frac{I_0\left(x,y\right)}{I_{sat}^*}(1-e^{-OD\left(x,y\right)}), \label{eq:imagingrewrite} 
\end{align}
where $I_t$ is the transmitted intensity after the atomic cloud, $I_0$ is the initial intensity before the atomic cloud, $I_{sat}^*$ is the effective saturation intensity, $\sigma_0^*$ is the effective scattering cross section and the optical density $OD$ is defined as $OD=-\ln\frac{I_t}{I_0}$.
 
Due to the uniform intensity distribution of the imaging beam at the position of the atoms, $I_0\left(x,y\right)$ is independent of $x$ and $y$ to a good approximation. In order to calibrate $I_0/I_{sat}^*$, we take several subsequent data sets of a pure atomic sample at $\unit[1400]{G}$ both with our regular imaging settings and with a $\unit[10]{dB}$ attenuated imaging intensity. We then use equation (\ref{eq:imagingrewrite}) and adjust $I_0/I_{sat}^*$ such that the RHS yields the same result both for the regular and the low-intensity setting. Averaging over the data sets then results in $I_0/I_{sat}^*=0.97_{-0.08}^{+0.13}$. The systematic uncertainties are estimated by the minimum and maximum $I_0/I_{sat}^*$ obtained for the individual data sets. This leads to a systematic uncertainty of $_{-4\%}^{+7\%}$ for the atom number $N$ and the peak density $n_{0}$, and a negligible uncertainty for $T$ and $N_q/N$. In addition, we independently measure the power of the imaging beam and thus determine the imaging intensity to be $I \approx I_{sat}$. This justifies using the literature value $\sigma_0$ \cite{gehm2003} for $\sigma_0^*$.

On the BEC-side, the binding energy of the molecules shifts the resonance frequency which leads to a decreased detection efficiency. Using in situ images at different fields, we calibrate this factor for our imaging settings. For magnetic fields below $782\,$G, it deviates from $1$ and reaches $N_{at}/N_{mol}=1.33_{-0.07}^{+0.10}$ at $692\,$G. This leads to a systematic uncertainty of up to $8$\% in atom number $N$ and peak density $n_{0}$ for the affected magnetic fields. More details about the systematic uncertainties can be found in section \ref{subsec:systematic_errors}.

The duration of our imaging pulse is $\tau = \unit[8]{\mu s}$. Due to the small mass of $^6$Li, the atoms are accelerated during the imaging pulse. This results in a Doppler shift of approximately $\unit[10]{MHz}$ at the end of the imaging pulse. To compensate for this effect, we linearly sweep the imaging laser frequency during the pulse. 
In order to reduce the shot noise in the absorption images, we use a ten times longer reference pulse. To further improve the quality of the absorption images, we apply a fringe removal algorithm \cite{Ockeloen2010}.

\section{Temperature determination}
\label{subsec:temp_det}

We obtain the temperature $T$ of each sample by fitting a Boltzmann distribution given by 
\begin{align}
  \tilde{n}(p,t=0) = n(x,t_{\text{exp}}) = A_0 \exp\left(- \frac{M \omega^2_{\text{exp}} x^2}{2 k_B T} \right) 
  \label{eq:boltzmann}
\end{align}
to the wing of the radial momentum distribution \cite{Murthy2014}. Here, $M$ is the mass of the expanding particle, $k_B$ is Boltzmann's constant, $A_0$ is the amplitude of the fit function, and $\omega_{\text{exp}}$ is the trapping frequency of the radial magnetic confinement the particles expand in. As evident from Fig. 3A, this function describes the data well over a range of more than $50$ pixels. The temperatures used in the main text are the average of approximately $30$ realizations. In order to obtain the degeneracy temperature $T/T_F$, we obtain $T_F$ from the in situ peak density of the sample as described in the main text.

For magnetic fields $\leq 782\,$G, the thermal part of the sample has the momentum distribution of molecules. We verify this in a measurement where we prepare the sample at different magnetic fields and let it evolve in time-of-flight for $3\,$ms before ramping the magnetic field to $527\,$G, where molecules are deeply bound and are thus not detected in absorption imaging resonant with free atoms. We observe that all atoms are bound in molecules after this expansion experiment at all investigated temperatures for magnetic fields $\leq 782\,$G. Thus, they are also bound in the trap. For these magnetic fields, 
we thus use the molecule mass in equation (\ref{eq:boltzmann}). For magnetic fields $\geq 892\,$G, the binding energy of the quasi-2D dimer is significantly smaller than the thermal energy in our sample.
The thermal wing thus has the momentum distribution of atoms and we use the atom mass. For intermediate fields, the thermal part crosses over from the molecular to the atomic momentum distribution. Thus, using the atom and molecule mass one obtains an upper and lower bound on the temperature.

\begin{figure} [!htb]
\centering
	\includegraphics [width= 7 cm] {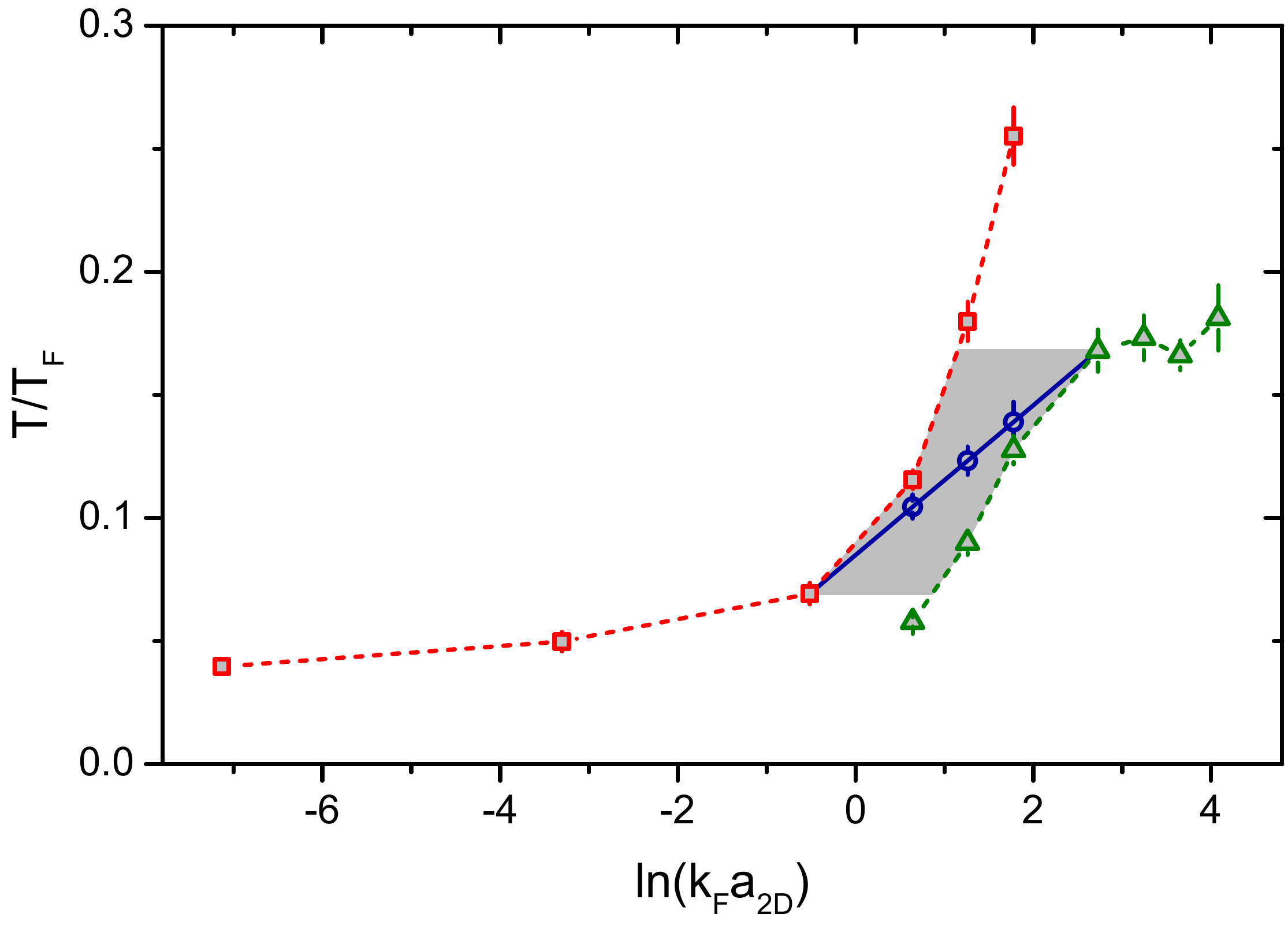}
	\caption{\textbf{Temperature interpolation in the strongly interacting regime} at the lowest attainable temperature. We obtain the temperature at magnetic fields $782\,\mathrm{G} < B < 892\,\mathrm{G}$, where the thermal part of the gas does not exclusively consist of molecules or atoms, from a linear interpolation between the points at $782\,$G and $892\,$G (solid line). The temperatures obtained with the molecule (atom) mass are depicted as red squares (green triangles). The systematic uncertainty of the interpolated temperatures are indicated by the gray area. Dashed lines are guides to the eye. Each data point is the average of approximately $30$ individual measurements, the error bars denote the standard error of the mean.}
	\label{fig:figures3}
\end{figure}

We determine the degeneracy temperature at the intermediate fields from a linear interpolation of $T/T_F$ versus $\ln(k_Fa_{2D})$ between $782\,$G and $892\,$G for samples where we applied the same heating parameter. This interpolation is depicted in Fig. S\ref{fig:figures3} for the lowest attainable temperature. The behavior of the interpolated temperature and of the temperatures obtained using the molecular (red) and atomic (green) mass justifies the interpolation procedure. The interpolated temperature always lies between the molecular and the atomic limit. It is close to the molecular limit on the BEC side, and crosses over to the atomic limit as $\ln(k_Fa_{2D})$ increases. 
We estimate the systematic error of the interpolated temperature using two assumptions: $T/T_F$ has to be monotonous in $\ln(k_Fa_{2D})$, which yields an interval between the $T/T_F$ of the two points between which we interpolate, and $T/T_F$ has to lie between the values obtained from a fit with the molecule mass and the atom mass (red and green data in Fig. S\ref{fig:figures3}). The overlap of these two intervals is indicated by the gray area in Fig. S\ref{fig:figures3}. It gives an upper bound for the systematic uncertainty of the interpolation result. The statistical errors of the interpolated $T/T_F$ are obtained from the statistical errors at $782\,$G and $892\,$G.

In addition to the temperature determination from the momentum distribution, we also extract the temperature from the in situ data. Applying the local density approximation to the whole cloud, we plot the in situ density as a function of the trapping potential $V(r)$ and fit its wing with a Boltzmann distribution, which in this case is given by 
$n(V) = B_0 \exp\left(- \frac{\alpha V}{k_B T} \right)$, where $1 \leq \alpha \leq 2$ takes into account whether the thermal wing consists of atoms or molecules. For the intermediate magnetic fields, we perform an interpolation similar to the one mentioned above to determine $\alpha$.

For magnetic fields up to $812\,$G, the temperatures obtained from both methods agree for low temperatures. For low fields and the highest temperatures, the in situ fit yields larger temperatures. For larger fields, the temperature from the in situ data is systematically lower than the temperature obtained from the momentum distribution. 
The reasons for this deviation are still unclear.
Nevertheless, within their errors, the extracted critical temperatures from both methods are still compatible with each other. The values for $T_c/T_F$ obtained with both methods are listed in table \ref{tab:tc}.

\begin{table*} [htb]
\begin{center}
\small
  \begin{tabular}{|c| c | c | c |c | c |} \hline
    \: $B$ [Gauss] \: &\: $\ell_z/a_{3D}$ \:&\: $\ln(k_Fa_{2D})_{T_c}$ (stat.)(sys.)\:& \: $\ln(k_F\tilde{a}_{2D})_{T_c}$ (stat.)(sys.)\:&\: $T_c/T_F$ (stat.)(sys.) \:&\: $(T_c/T_F)_{\text{in situ}}$ (stat.)\:   \\ 
    \hline 
    & & & & & \\
    692	& \,\,7.11 	&-\,7.30 \,(4) $\left(_{-5}^{+4}\right)$		&-\,0.96 \,(4)$\left(_{-3}^{+2}\right)$			& 0.089 \, (15)  $\left(_{-13}^{+14}\right)$	& 0.090 \, (13)  	\\ [5pt]
	732	& \,\,3.98 	&-\,3.42 \,(2) $\left(_{-6}^{+4}\right)$		&-\,0.45 \,(2)$\left(_{-5}^{+3}\right)$			& 0.100 \, (22)  $\left(_{-15}^{+17}\right)$	& 0.099 \, (27)   	\\ [5pt]
    782	& \,\,1.55	&-\,0.59 \,(1) $\left(_{-7}^{+4}\right)$		&\,\,\,0.20 \,(1)$\left(_{-6}^{+4}\right)$		& 0.129 \, (35)  $\left(_{-18}^{+24}\right)$	& 0.112 \, (44) 	\\ [5pt]
	812	& \,\,0.55	&\,\,\,0.57 \,(1) $\left(_{-7}^{+2}\right)$	&\,\,\,0.79 \,(2)$\left(_{-7}^{+2}\right)$		& 0.146 \, (25)  $\left(_{-23}^{+50}\right)$	& 0.146 \, (21)  	\\ [5pt]
	832	& \,\,0		&\,\,\,1.23 \,(1) $\left(_{-8}^{+2}\right)$	&\,\,\,1.33 \,(1)$\left(_{-8}^{+2}\right)$		& 0.167 \, (39)  $\left( _{-34}^{+48}\right)$	& 0.122   (103)  	\\ [5pt]
	852	& -0.46		&\,\,\,1.72 \,(1) $\left(_{-9}^{+2}\right)$  &\,\,\,1.76 \,(1)$\left(_{-9}^{+2}\right)$		& 0.167 \, (27)  $\left( _{-22}^{+42}\right)$	& 0.122 \, (60)  	\\ [5pt]
  \hline 
  \end{tabular}
\caption{\textbf{Measured critical temperatures.}
The measured critical temperatures $T_c/T_F$ are given with their respective statistical and systematic errors as a function of the magnetic offset field $B$, $\ell_z/a_{3D}$, and the 2D interaction parameter $\ln(k_Fa_{2D})$ at $T_c/T_F$. In addition, also the alternative 2D interaction parameter $\ln(k_F\tilde{a}_{2D})_{T_c}$ obtained with equation (\ref{EqTrans}) and the critical temperature $(T_c/T_F)_{\text{in situ}}$ obtained from the in situ temperature fit are given.}
  \label{tab:tc}
\end{center}
\vskip -0.5cm
\end {table*}

\section{Systematic uncertainties}
\label{subsec:systematic_errors}

The errors given in the main text are the statistical errors of our measurements. In addition, systematic uncertainties arise due to uncertainties in the following quantities:

\textbf{Imaging intensity $\boldsymbol{I_0/I_{sat}}$:} The measured atom density depends on the intensity $I_0$ of the imaging beam (see equation (\ref{eq:imagingrewrite})). In our experiments we use an imaging intensity of $I_0/I_{sat}^*=0.97_{-0.08}^{+0.13}$ (see section \ref{subsec:imaging}). This leads to an uncertainty in the peak density $n_{0}$ and thus we obtain $T_F$ $_{-4\%}^{+7\%}$ and $k_F$ $_{-2\%}^{+3.5\%}$.

\textbf{Atoms in non-central pancakes:} We obtain an upper bound of $11\%$ for the fraction of atoms in the non-central layer of the SWT (see section \ref{subsec:tomography}). To estimate their influence on the measured peak density $n_{0}$, we assume that their temperature is not affected by the evaporative cooling in the SWT because of the small atom number. However, it is affected by the heating procedure where the depth of the trap is modulated. The minimum temperature of these atoms is thus about $100\,$nK, the temperature at the transfer into the SWT. Assuming a thermal Boltzmann gas, we calculate the density of $5500$ atoms in the two non-central layers to be $n_{0,\text{nonc.}} = 0.14\,$atoms$/\mu$m$^2$ \cite{Ketterle2008}. This leads to an overestimation of $T_F$ by $5\%$ for the lowest magnetic fields, and $19\%$ at the highest magnetic fields, where $n_{0}$ is smaller due to the fermionic character of the sample. Analogously, $k_F$ is overestimated by $2.5\%$ to $9.5\%$.

\textbf{Reduced absorption cross section of molecules:} A finite molecular binding energy leads to a shift of the optical transition frequency of the dimers and results in a reduced absorption cross section. By rescaling the obtained images for the lowest three magnetic fields ($692\,$G, $732\,$G and $782\,$G), we compensate for this effect (see section \ref{subsec:imaging}). The uncertainty in the rescaling factor leads to an uncertainty in the peak density $n_{0}$ and thus $T_F$ which is smaller than $8\%$. The corresponding uncertainty in $k_F$ is smaller than $4\%$.

\textbf{Magnification of the imaging system:} We calibrate the magnification of the imaging system using Kapitza-Dirac scattering of atoms on an optical standing wave potential with known periodicity. The resulting uncertainty in the magnification is approximately $3\%$. This uncertainty quadratically enters the temperature obtained from the fit with equation (\ref{eq:boltzmann}). Thus, $T$ and $T/T_F$ have a relative systematic uncertainty of approximately $6\%$. 

\textbf{Trap frequency of the expansion potential:} We measure the trap frequency of the expansion potential $\omega_{\text{exp}}$ with a relative uncertainty of approximately $3\%$. This leads to an uncertainty of $6\%$ in $T$ (see equation (\ref*{eq:boltzmann})).

\textbf{Expansion time:} The measured momentum distribution depends on the expansion time. The extracted temperature has a maximum for $t_{\text{exp}}=T/4=\pi/(2\omega_{\text{exp}})$. Deviations from the ideal expansion time lead to a systematic underestimation of $T$ by approximately $5$\%.

\textbf{Fitted region for temperature determination:} The temperature determined from the fit to the tail of the  momentum distribution weakly depends on the region included in the fit. This leads to a relative uncertainty of $7$\% in $T$ for low temperatures and up to $13$\% at the highest investigated temperatures. The same effect also leads to a systematic uncertainty in the determined nonthermal fraction $N_q/N$. The absolute uncertainty in $N_q/N$ ranges from $0.06$ at low temperatures to about $0.02$ at high temperatures. 

\textbf{Temperature interpolation ($\boldsymbol{812}\,$G - $\boldsymbol{852}\,$G):} For the magnetic fields $812\,$G, $832\,$G and  $852\,$G, the reduced temperature $T/T_F$ is determined by interpolation because here the sample consists of a mixture of molecules and atoms (see section \ref{subsec:temp_det}). The systematic uncertainties are estimated as detailed in section \ref{subsec:temp_det}.
The obtained relative uncertainties are largest for low temperatures. They are usually on the order of $10$\%-$25$\% and range up to $50$\% for few individual values. For the critical temperature $T_c$, the total systematic uncertainties including those from temperature interpolation are listed in table \ref{tab:tc}.

\textbf{Resulting total systematic uncertainties:} Assuming a Gaussian distribution of the previously mentioned $8$ independent error sources, we calculate the total relative systematic uncertainty of $T/T_F$. 
For magnetic fields where the temperature is not interpolated, one obtains $T/T_F$$_{-13\%}^{+15\%}$ for low temperatures and low magnetic fields to $T/T_F$$_{-15\%}^{+28\%}$ at high temperatures and high magnetic fields. At the critical temperature, one obtains approximately $T_c/T_F$$_{-15\%}^{+17\%}$.
At the magnetic fields where the temperature is determined by interpolation, the systematic uncertainties are significantly bigger and one obtains approximately $T_c/T_F$$_{-20\%}^{+30\%}$ at the critical temperature. 
The systematic uncertainty in $\ln(k_F a_{2D})$ is dominated by the uncertainties in $k_F$. We thus neglect uncertainties in $a_{2D}$. The total absolute systematic uncertainty then lies between $\ln(k_F a_{2D})$$_{-0.05}^{+0.04}$ for low magnetic fields and $\ln(k_F a_{2D})$$_{-0.10}^{+0.02}$ for high magnetic fields.
The systematic uncertainties of $T/T_F$ and $\ln(k_F a_{2D})$ at the critical temperature $T_c$ are listed in table \ref{tab:tc}.

\textbf{Temperature determination from in situ profiles:} As mentioned in section \ref{subsec:temp_det}, we additionally determine the temperature from the in situ profiles for comparison. For magnetic fields $\ge 832\,$G, these temperatures are systematically smaller than those determined from the momentum distribution. The corresponding values for $(T_c/T_F)_{\text{in situ}}$ are listed in table \ref{tab:tc}.


\begin{thebibliography}{53}
\expandafter\ifx\csname natexlab\endcsname\relax\def\natexlab#1{#1}\fi
\expandafter\ifx\csname bibnamefont\endcsname\relax
  \def\bibnamefont#1{#1}\fi
\expandafter\ifx\csname bibfnamefont\endcsname\relax
  \def\bibfnamefont#1{#1}\fi
\expandafter\ifx\csname citenamefont\endcsname\relax
  \def\citenamefont#1{#1}\fi
\expandafter\ifx\csname url\endcsname\relax
  \def\url#1{\texttt{#1}}\fi
\expandafter\ifx\csname urlprefix\endcsname\relax\def\urlprefix{URL }\fi
\providecommand{\bibinfo}[2]{#2}
\providecommand{\eprint}[2][]{\url{#2}}

\bibitem[{\citenamefont{Norman}(2011)}]{Norman2011}
\bibinfo{author}{\bibfnamefont{M.~R.} \bibnamefont{Norman}},
  \bibinfo{journal}{Science} \textbf{\bibinfo{volume}{332}},
  \bibinfo{pages}{196} (\bibinfo{year}{2011}),
  \urlprefix\url{http://dx.doi.org/10.1126/science.1200181}.

\bibitem[{\citenamefont{Lee et~al.}(2006)\citenamefont{Lee, Nagaosa, and
  Wen}}]{Lee2006}
\bibinfo{author}{\bibfnamefont{P.~A.} \bibnamefont{Lee}},
  \bibinfo{author}{\bibfnamefont{N.}~\bibnamefont{Nagaosa}}, \bibnamefont{and}
  \bibinfo{author}{\bibfnamefont{X.-G.} \bibnamefont{Wen}},
  \bibinfo{journal}{Rev. Mod. Phys.} \textbf{\bibinfo{volume}{78}},
  \bibinfo{pages}{17} (\bibinfo{year}{2006}),
  \urlprefix\url{http://link.aps.org/doi/10.1103/RevModPhys.78.17}.

\bibitem[{\citenamefont{{Levinsen} and {Parish}}(2014)}]{Levinsen2014}
\bibinfo{author}{\bibfnamefont{J.}~\bibnamefont{{Levinsen}}} \bibnamefont{and}
  \bibinfo{author}{\bibfnamefont{M.~M.} \bibnamefont{{Parish}}},
  \bibinfo{journal}{ArXiv e-prints}  (\bibinfo{year}{2014}),
  \eprint{1408.2737}, \urlprefix\url{http://arxiv.org/abs/1408.2737}.

\bibitem[{\citenamefont{Randeria et~al.}(1989)\citenamefont{Randeria, Duan, and
  Shieh}}]{Randeria1989}
\bibinfo{author}{\bibfnamefont{M.}~\bibnamefont{Randeria}},
  \bibinfo{author}{\bibfnamefont{J.-M.} \bibnamefont{Duan}}, \bibnamefont{and}
  \bibinfo{author}{\bibfnamefont{L.-Y.} \bibnamefont{Shieh}},
  \bibinfo{journal}{Phys. Rev. Lett.} \textbf{\bibinfo{volume}{62}},
  \bibinfo{pages}{981} (\bibinfo{year}{1989}),
  \urlprefix\url{http://link.aps.org/doi/10.1103/PhysRevLett.62.981}.

\bibitem[{\citenamefont{Iskin and de~Melo}(2009)}]{Iskin2009}
\bibinfo{author}{\bibfnamefont{M.}~\bibnamefont{Iskin}} \bibnamefont{and}
  \bibinfo{author}{\bibfnamefont{C.~A. R.~S.} \bibnamefont{de~Melo}},
  \bibinfo{journal}{Phys. Rev. Lett.} \textbf{\bibinfo{volume}{103}},
  \bibinfo{pages}{165301} (\bibinfo{year}{2009}),
  \urlprefix\url{http://link.aps.org/doi/10.1103/PhysRevLett.103.165301}.

\bibitem[{\citenamefont{Bertaina and Giorgini}(2011)}]{Bertaina2011}
\bibinfo{author}{\bibfnamefont{G.}~\bibnamefont{Bertaina}} \bibnamefont{and}
  \bibinfo{author}{\bibfnamefont{S.}~\bibnamefont{Giorgini}},
  \bibinfo{journal}{Phys. Rev. Lett.} \textbf{\bibinfo{volume}{106}},
  \bibinfo{pages}{110403} (\bibinfo{year}{2011}),
  \urlprefix\url{http://link.aps.org/doi/10.1103/PhysRevLett.106.110403}.

\bibitem[{\citenamefont{Bauer et~al.}(2014)\citenamefont{Bauer, Parish, and
  Enss}}]{Bauer2014}
\bibinfo{author}{\bibfnamefont{M.}~\bibnamefont{Bauer}},
  \bibinfo{author}{\bibfnamefont{M.~M.} \bibnamefont{Parish}},
  \bibnamefont{and} \bibinfo{author}{\bibfnamefont{T.}~\bibnamefont{Enss}},
  \bibinfo{journal}{Phys. Rev. Lett.} \textbf{\bibinfo{volume}{112}},
  \bibinfo{pages}{135302} (\bibinfo{year}{2014}),
  \urlprefix\url{http://link.aps.org/doi/10.1103/PhysRevLett.112.135302}.

\bibitem[{\citenamefont{{Matsumoto} and {Ohashi}}(2014)}]{Matsumoto2014}
\bibinfo{author}{\bibfnamefont{M.}~\bibnamefont{{Matsumoto}}} \bibnamefont{and}
  \bibinfo{author}{\bibfnamefont{Y.}~\bibnamefont{{Ohashi}}},
  \bibinfo{journal}{J. Phys.: Conf. Ser.} \textbf{\bibinfo{volume}{568}},
    \bibinfo{pages}{012012} (\bibinfo{year}{2014}), 
    \urlprefix\url{http://iopscience.iop.org/1742-6596/568/1/012012/pdf/1742-6596_568_1_012012.pdf}.

\bibitem[{\citenamefont{Bloch et~al.}(2008)\citenamefont{Bloch, Dalibard, and
  Zwerger}}]{Bloch2008}
\bibinfo{author}{\bibfnamefont{I.}~\bibnamefont{Bloch}},
  \bibinfo{author}{\bibfnamefont{J.}~\bibnamefont{Dalibard}}, \bibnamefont{and}
  \bibinfo{author}{\bibfnamefont{W.}~\bibnamefont{Zwerger}},
  \bibinfo{journal}{Rev. Mod. Phys.} \textbf{\bibinfo{volume}{80}},
  \bibinfo{pages}{885} (\bibinfo{year}{2008}),
  \urlprefix\url{http://link.aps.org/doi/10.1103/RevModPhys.80.885}.

\bibitem[{\citenamefont{{H}adzibabic et~al.}(2006)\citenamefont{{H}adzibabic,
  {K}r{\"u}ger, {C}heneau, {B}attelier, and {D}alibard}}]{Hadzibabic2006}
\bibinfo{author}{\bibfnamefont{Z.}~\bibnamefont{{H}adzibabic}},
  \bibinfo{author}{\bibfnamefont{P.}~\bibnamefont{{K}r{\"u}ger}},
  \bibinfo{author}{\bibfnamefont{M.}~\bibnamefont{{C}heneau}},
  \bibinfo{author}{\bibfnamefont{B.}~\bibnamefont{{B}attelier}},
  \bibnamefont{and}
  \bibinfo{author}{\bibfnamefont{J.}~\bibnamefont{{D}alibard}},
  \bibinfo{journal}{{N}ature} \textbf{\bibinfo{volume}{441}},
  \bibinfo{pages}{1118} (\bibinfo{year}{2006}),
  \urlprefix\url{http://dx.doi.org/10.1038/nature04851}.

\bibitem[{\citenamefont{Desbuquois et~al.}(2012)\citenamefont{Desbuquois,
  Chomaz, Yefsah, Leonard, Beugnon, Weitenberg, and Dalibard}}]{Desbuquois2012}
\bibinfo{author}{\bibfnamefont{R.}~\bibnamefont{Desbuquois}},
  \bibinfo{author}{\bibfnamefont{L.}~\bibnamefont{Chomaz}},
  \bibinfo{author}{\bibfnamefont{T.}~\bibnamefont{Yefsah}},
  \bibinfo{author}{\bibfnamefont{J.}~\bibnamefont{Leonard}},
  \bibinfo{author}{\bibfnamefont{J.}~\bibnamefont{Beugnon}},
  \bibinfo{author}{\bibfnamefont{C.}~\bibnamefont{Weitenberg}},
  \bibnamefont{and} \bibinfo{author}{\bibfnamefont{J.}~\bibnamefont{Dalibard}},
  \bibinfo{journal}{Nature Physics} \textbf{\bibinfo{volume}{8}},
  \bibinfo{pages}{645} (\bibinfo{year}{2012}),
  \urlprefix\url{http://dx.doi.org/10.1038/nphys2378}.

\bibitem[{\citenamefont{Bartenstein et~al.}(2004)\citenamefont{Bartenstein,
  Altmeyer, Riedl, Jochim, Chin, Denschlag, and Grimm}}]{Bartenstein2004}
\bibinfo{author}{\bibfnamefont{M.}~\bibnamefont{Bartenstein}},
  \bibinfo{author}{\bibfnamefont{A.}~\bibnamefont{Altmeyer}},
  \bibinfo{author}{\bibfnamefont{S.}~\bibnamefont{Riedl}},
  \bibinfo{author}{\bibfnamefont{S.}~\bibnamefont{Jochim}},
  \bibinfo{author}{\bibfnamefont{C.}~\bibnamefont{Chin}},
  \bibinfo{author}{\bibfnamefont{J.~H.} \bibnamefont{Denschlag}},
  \bibnamefont{and} \bibinfo{author}{\bibfnamefont{R.}~\bibnamefont{Grimm}},
  \bibinfo{journal}{Phys. Rev. Lett.} \textbf{\bibinfo{volume}{92}},
  \bibinfo{pages}{203201} (\bibinfo{year}{2004}),
  \urlprefix\url{http://link.aps.org/doi/10.1103/PhysRevLett.92.203201}.

\bibitem[{\citenamefont{{R}egal et~al.}(2004)\citenamefont{{R}egal, {G}reiner,
  and {J}in}}]{Regal2004}
\bibinfo{author}{\bibfnamefont{C.~A.} \bibnamefont{{R}egal}},
  \bibinfo{author}{\bibfnamefont{M.}~\bibnamefont{{G}reiner}},
  \bibnamefont{and} \bibinfo{author}{\bibfnamefont{D.~S.} \bibnamefont{{J}in}},
  \bibinfo{journal}{{P}hys. {R}ev. {L}ett.} \textbf{\bibinfo{volume}{92}},
  \bibinfo{pages}{040403} (\bibinfo{year}{2004}),
  \urlprefix\url{http://link.aps.org/abstract/PRL/v92/e040403}.

\bibitem[{\citenamefont{Zwierlein et~al.}(2004)\citenamefont{Zwierlein, Stan,
  Schunck, Raupach, Kerman, and Ketterle}}]{Zwierlein2004}
\bibinfo{author}{\bibfnamefont{M.~W.} \bibnamefont{Zwierlein}},
  \bibinfo{author}{\bibfnamefont{C.~A.} \bibnamefont{Stan}},
  \bibinfo{author}{\bibfnamefont{C.~H.} \bibnamefont{Schunck}},
  \bibinfo{author}{\bibfnamefont{S.~M.~F.} \bibnamefont{Raupach}},
  \bibinfo{author}{\bibfnamefont{A.~J.} \bibnamefont{Kerman}},
  \bibnamefont{and} \bibinfo{author}{\bibfnamefont{W.}~\bibnamefont{Ketterle}},
  \bibinfo{journal}{Phys. Rev. Lett.} \textbf{\bibinfo{volume}{92}},
  \bibinfo{pages}{120403} (\bibinfo{year}{2004}),
  \urlprefix\url{http://link.aps.org/doi/10.1103/PhysRevLett.92.120403}.

\bibitem[{\citenamefont{Bourdel et~al.}(2004)\citenamefont{Bourdel, Khaykovich,
  Cubizolles, Zhang, Chevy, Teichmann, Tarruell, Kokkelmans, and
  Salomon}}]{Bourdel2004}
\bibinfo{author}{\bibfnamefont{T.}~\bibnamefont{Bourdel}},
  \bibinfo{author}{\bibfnamefont{L.}~\bibnamefont{Khaykovich}},
  \bibinfo{author}{\bibfnamefont{J.}~\bibnamefont{Cubizolles}},
  \bibinfo{author}{\bibfnamefont{J.}~\bibnamefont{Zhang}},
  \bibinfo{author}{\bibfnamefont{F.}~\bibnamefont{Chevy}},
  \bibinfo{author}{\bibfnamefont{M.}~\bibnamefont{Teichmann}},
  \bibinfo{author}{\bibfnamefont{L.}~\bibnamefont{Tarruell}},
  \bibinfo{author}{\bibfnamefont{S.~J. J. M.~F.} \bibnamefont{Kokkelmans}},
  \bibnamefont{and} \bibinfo{author}{\bibfnamefont{C.}~\bibnamefont{Salomon}},
  \bibinfo{journal}{Phys. Rev. Lett.} \textbf{\bibinfo{volume}{93}},
  \bibinfo{pages}{050401} (\bibinfo{year}{2004}),
  \urlprefix\url{http://link.aps.org/doi/10.1103/PhysRevLett.93.050401}.

\bibitem[{\citenamefont{Martiyanov et~al.}(2010)\citenamefont{Martiyanov,
  Makhalov, and Turlapov}}]{Martiyanov2010}
\bibinfo{author}{\bibfnamefont{K.}~\bibnamefont{Martiyanov}},
  \bibinfo{author}{\bibfnamefont{V.}~\bibnamefont{Makhalov}}, \bibnamefont{and}
  \bibinfo{author}{\bibfnamefont{A.}~\bibnamefont{Turlapov}},
  \bibinfo{journal}{Phys. Rev. Lett.} \textbf{\bibinfo{volume}{105}},
  \bibinfo{pages}{030404} (\bibinfo{year}{2010}),
  \urlprefix\url{http://link.aps.org/doi/10.1103/PhysRevLett.105.030404}.

\bibitem[{\citenamefont{Feld et~al.}(2011)\citenamefont{Feld, Fr\"ohlich, Vogt,
  Koschorreck, and K\"ohl}}]{Feld2011}
\bibinfo{author}{\bibfnamefont{M.}~\bibnamefont{Feld}},
  \bibinfo{author}{\bibfnamefont{B.}~\bibnamefont{Fr\"ohlich}},
  \bibinfo{author}{\bibfnamefont{E.}~\bibnamefont{Vogt}},
  \bibinfo{author}{\bibfnamefont{M.}~\bibnamefont{Koschorreck}},
  \bibnamefont{and} \bibinfo{author}{\bibfnamefont{M.}~\bibnamefont{K\"ohl}},
  \bibinfo{journal}{Nature} \textbf{\bibinfo{volume}{480}}, \bibinfo{pages}{75}
  (\bibinfo{year}{2011}),
  \urlprefix\url{http://dx.doi.org/10.1038/nature10627}.

\bibitem[{\citenamefont{Dyke et~al.}(2011)\citenamefont{Dyke, Kuhnle, Whitlock,
  Hu, Mark, Hoinka, Lingham, Hannaford, and Vale}}]{Dyke2011}
\bibinfo{author}{\bibfnamefont{P.}~\bibnamefont{Dyke}},
  \bibinfo{author}{\bibfnamefont{E.~D.} \bibnamefont{Kuhnle}},
  \bibinfo{author}{\bibfnamefont{S.}~\bibnamefont{Whitlock}},
  \bibinfo{author}{\bibfnamefont{H.}~\bibnamefont{Hu}},
  \bibinfo{author}{\bibfnamefont{M.}~\bibnamefont{Mark}},
  \bibinfo{author}{\bibfnamefont{S.}~\bibnamefont{Hoinka}},
  \bibinfo{author}{\bibfnamefont{M.}~\bibnamefont{Lingham}},
  \bibinfo{author}{\bibfnamefont{P.}~\bibnamefont{Hannaford}},
  \bibnamefont{and} \bibinfo{author}{\bibfnamefont{C.~J.} \bibnamefont{Vale}},
  \bibinfo{journal}{Phys. Rev. Lett.} \textbf{\bibinfo{volume}{106}},
  \bibinfo{pages}{105304} (\bibinfo{year}{2011}),
  \urlprefix\url{http://link.aps.org/doi/10.1103/PhysRevLett.106.105304}.

\bibitem[{\citenamefont{Koschorreck et~al.}(2012)\citenamefont{Koschorreck,
  Pertot, Vogt, Fr\"ohlich, Feld, and K\"ohl}}]{Koschorreck2012}
\bibinfo{author}{\bibfnamefont{M.}~\bibnamefont{Koschorreck}},
  \bibinfo{author}{\bibfnamefont{D.}~\bibnamefont{Pertot}},
  \bibinfo{author}{\bibfnamefont{E.}~\bibnamefont{Vogt}},
  \bibinfo{author}{\bibfnamefont{B.}~\bibnamefont{Fr\"ohlich}},
  \bibinfo{author}{\bibfnamefont{M.}~\bibnamefont{Feld}}, \bibnamefont{and}
  \bibinfo{author}{\bibfnamefont{M.}~\bibnamefont{K\"ohl}},
  \bibinfo{journal}{Nature} \textbf{\bibinfo{volume}{485}},
  \bibinfo{pages}{619} (\bibinfo{year}{2012}),
  \urlprefix\url{http://dx.doi.org/10.1038/nature11151}.

\bibitem[{\citenamefont{Sommer et~al.}(2012)\citenamefont{Sommer, Cheuk, Ku,
  Bakr, and Zwierlein}}]{Sommer2012}
\bibinfo{author}{\bibfnamefont{A.~T.} \bibnamefont{Sommer}},
  \bibinfo{author}{\bibfnamefont{L.~W.} \bibnamefont{Cheuk}},
  \bibinfo{author}{\bibfnamefont{M.~J.~H.} \bibnamefont{Ku}},
  \bibinfo{author}{\bibfnamefont{W.~S.} \bibnamefont{Bakr}}, \bibnamefont{and}
  \bibinfo{author}{\bibfnamefont{M.~W.} \bibnamefont{Zwierlein}},
  \bibinfo{journal}{Phys. Rev. Lett.} \textbf{\bibinfo{volume}{108}},
  \bibinfo{pages}{045302} (\bibinfo{year}{2012}),
  \urlprefix\url{http://link.aps.org/doi/10.1103/PhysRevLett.108.045302}.

\bibitem[{\citenamefont{Makhalov et~al.}(2014)\citenamefont{Makhalov,
  Martiyanov, and Turlapov}}]{Makhalov2014}
\bibinfo{author}{\bibfnamefont{V.}~\bibnamefont{Makhalov}},
  \bibinfo{author}{\bibfnamefont{K.}~\bibnamefont{Martiyanov}},
  \bibnamefont{and} \bibinfo{author}{\bibfnamefont{A.}~\bibnamefont{Turlapov}},
  \bibinfo{journal}{Phys. Rev. Lett.} \textbf{\bibinfo{volume}{112}},
  \bibinfo{pages}{045301} (\bibinfo{year}{2014}),
  \urlprefix\url{http://link.aps.org/doi/10.1103/PhysRevLett.112.045301}.

\bibitem[{\citenamefont{Mermin and Wagner}(1966)}]{Mermin1966}
\bibinfo{author}{\bibfnamefont{N.~D.} \bibnamefont{Mermin}} \bibnamefont{and}
  \bibinfo{author}{\bibfnamefont{H.}~\bibnamefont{Wagner}},
  \bibinfo{journal}{Phys. Rev. Lett.} \textbf{\bibinfo{volume}{17}},
  \bibinfo{pages}{113} (\bibinfo{year}{1966}),
  \urlprefix\url{http://link.aps.org/doi/10.1103/PhysRevLett.17.1133}.

\bibitem[{\citenamefont{Hohenberg}(1967)}]{Hohenberg1967}
\bibinfo{author}{\bibfnamefont{P.~C.} \bibnamefont{Hohenberg}},
  \bibinfo{journal}{Phys. Rev.} \textbf{\bibinfo{volume}{158}},
  \bibinfo{pages}{383} (\bibinfo{year}{1967}),
  \urlprefix\url{http://link.aps.org/doi/10.1103/PhysRev.158.383}.

\bibitem[{\citenamefont{Berezinskii}(1972)}]{Berezinskii1972}
\bibinfo{author}{\bibfnamefont{V.~L.} \bibnamefont{Berezinskii}},
  \bibinfo{journal}{Sov. Phys. JETP} \textbf{\bibinfo{volume}{34}},
  \bibinfo{pages}{610} (\bibinfo{year}{1972}).

\bibitem[{\citenamefont{Kosterlitz and Thouless}(1973)}]{Kosterlitz1973}
\bibinfo{author}{\bibfnamefont{J.~M.} \bibnamefont{Kosterlitz}}
  \bibnamefont{and} \bibinfo{author}{\bibfnamefont{D.~J.}
  \bibnamefont{Thouless}}, \bibinfo{journal}{Journal of Physics C: Solid State
  Physics} \textbf{\bibinfo{volume}{6}}, \bibinfo{pages}{1181}
  (\bibinfo{year}{1973}),
  \urlprefix\url{http://stacks.iop.org/0022-3719/6/i=7/a=010}.

\bibitem[{\citenamefont{Z\"urn et~al.}(2013)\citenamefont{Z\"urn, Lompe, Wenz,
  Jochim, Julienne, and Hutson}}]{Zuern2013}
\bibinfo{author}{\bibfnamefont{G.}~\bibnamefont{Z\"urn}},
  \bibinfo{author}{\bibfnamefont{T.}~\bibnamefont{Lompe}},
  \bibinfo{author}{\bibfnamefont{A.~N.} \bibnamefont{Wenz}},
  \bibinfo{author}{\bibfnamefont{S.}~\bibnamefont{Jochim}},
  \bibinfo{author}{\bibfnamefont{P.~S.} \bibnamefont{Julienne}},
  \bibnamefont{and} \bibinfo{author}{\bibfnamefont{J.~M.}
  \bibnamefont{Hutson}}, \bibinfo{journal}{Phys. Rev. Lett.}
  \textbf{\bibinfo{volume}{110}}, \bibinfo{pages}{135301}
  (\bibinfo{year}{2013}),
  \urlprefix\url{http://link.aps.org/doi/10.1103/PhysRevLett.110.135301}.

\bibitem[{\citenamefont{See Supplemental Material [URL]}()}]{SOM}
\bibinfo{author}{\bibnamefont{See}} \bibnamefont{Supplemental}
  \bibinfo{author}{\bibnamefont{Material, \urlprefix\url{http://prl.aps.org/XXX}}}.

\bibitem[{\citenamefont{Petrov et~al.}(2000)\citenamefont{Petrov, Holzmann, and
  Shlyapnikov}}]{petrov2000b}
\bibinfo{author}{\bibfnamefont{D.~S.} \bibnamefont{Petrov}},
  \bibinfo{author}{\bibfnamefont{M.}~\bibnamefont{Holzmann}}, \bibnamefont{and}
  \bibinfo{author}{\bibfnamefont{G.~V.} \bibnamefont{Shlyapnikov}},
  \bibinfo{journal}{Phys. Rev. Lett.} \textbf{\bibinfo{volume}{84}},
  \bibinfo{pages}{2551} (\bibinfo{year}{2000}),
  \urlprefix\url{http://link.aps.org/doi/10.1103/PhysRevLett.84.2551}.

\bibitem[{\citenamefont{{S}hvarchuck et~al.}(2002)\citenamefont{{S}hvarchuck,
  {B}uggle, {P}etrov, {D}ieckmann, {Z}ielonkowski, {K}emmann, {T}iecke, von
  {K}litzing, {S}hlyapnikov, and {W}alraven}}]{Shvarchuck2002}
\bibinfo{author}{\bibfnamefont{I.}~\bibnamefont{{S}hvarchuck}},
  \bibinfo{author}{\bibfnamefont{C.}~\bibnamefont{{B}uggle}},
  \bibinfo{author}{\bibfnamefont{D.~S.} \bibnamefont{{P}etrov}},
  \bibinfo{author}{\bibfnamefont{K.}~\bibnamefont{{D}ieckmann}},
  \bibinfo{author}{\bibfnamefont{M.}~\bibnamefont{{Z}ielonkowski}},
  \bibinfo{author}{\bibfnamefont{M.}~\bibnamefont{{K}emmann}},
  \bibinfo{author}{\bibfnamefont{T.~G.} \bibnamefont{{T}iecke}},
  \bibinfo{author}{\bibfnamefont{W.}~\bibnamefont{von {K}litzing}},
  \bibinfo{author}{\bibfnamefont{G.~V.} \bibnamefont{{S}hlyapnikov}},
  \bibnamefont{and} \bibinfo{author}{\bibfnamefont{J.~T.~M.}
  \bibnamefont{{W}alraven}}, \bibinfo{journal}{{P}hys. {R}ev. {L}ett.}
  \textbf{\bibinfo{volume}{89}}, \bibinfo{pages}{270404}
  (\bibinfo{year}{2002}),
  \urlprefix\url{http://link.aps.org/abstract/PRL/v89/e270404}.

\bibitem[{\citenamefont{van Amerongen et~al.}(2008)\citenamefont{van Amerongen,
  van Es, Wicke, Kheruntsyan, and van Druten}}]{VanAmerongen2008}
\bibinfo{author}{\bibfnamefont{A.~H.} \bibnamefont{van Amerongen}},
  \bibinfo{author}{\bibfnamefont{J.~J.~P.} \bibnamefont{van Es}},
  \bibinfo{author}{\bibfnamefont{P.}~\bibnamefont{Wicke}},
  \bibinfo{author}{\bibfnamefont{K.~V.} \bibnamefont{Kheruntsyan}},
  \bibnamefont{and} \bibinfo{author}{\bibfnamefont{N.~J.} \bibnamefont{van
  Druten}}, \bibinfo{journal}{Phys. Rev. Lett.} \textbf{\bibinfo{volume}{100}}
  (\bibinfo{year}{2008}),
  \urlprefix\url{http://link.aps.org/doi/10.1103/PhysRevLett.100.090402}.

\bibitem[{\citenamefont{Tung et~al.}(2010)\citenamefont{Tung, Lamporesi,
  Lobser, Xia, and Cornell}}]{Tung2010}
\bibinfo{author}{\bibfnamefont{S.}~\bibnamefont{Tung}},
  \bibinfo{author}{\bibfnamefont{G.}~\bibnamefont{Lamporesi}},
  \bibinfo{author}{\bibfnamefont{D.}~\bibnamefont{Lobser}},
  \bibinfo{author}{\bibfnamefont{L.}~\bibnamefont{Xia}}, \bibnamefont{and}
  \bibinfo{author}{\bibfnamefont{E.~A.} \bibnamefont{Cornell}},
  \bibinfo{journal}{Phys. Rev. Lett.} \textbf{\bibinfo{volume}{105}},
  \bibinfo{pages}{230408} (\bibinfo{year}{2010}),
  \urlprefix\url{http://link.aps.org/doi/10.1103/PhysRevLett.105.230408}.

\bibitem[{\citenamefont{Murthy et~al.}(2014)\citenamefont{Murthy, Kedar, Lompe,
  Neidig, Ries, Wenz, Z\"urn, and Jochim}}]{Murthy2014}
\bibinfo{author}{\bibfnamefont{P.~A.} \bibnamefont{Murthy}},
  \bibinfo{author}{\bibfnamefont{D.}~\bibnamefont{Kedar}},
  \bibinfo{author}{\bibfnamefont{T.}~\bibnamefont{Lompe}},
  \bibinfo{author}{\bibfnamefont{M.}~\bibnamefont{Neidig}},
  \bibinfo{author}{\bibfnamefont{M.~G.} \bibnamefont{Ries}},
  \bibinfo{author}{\bibfnamefont{A.~N.} \bibnamefont{Wenz}},
  \bibinfo{author}{\bibfnamefont{G.}~\bibnamefont{Z\"urn}}, \bibnamefont{and}
  \bibinfo{author}{\bibfnamefont{S.}~\bibnamefont{Jochim}},
  \bibinfo{journal}{Phys. Rev. A} \textbf{\bibinfo{volume}{90}},
  \bibinfo{pages}{043611} (\bibinfo{year}{2014}),
  \urlprefix\url{http://link.aps.org/doi/10.1103/PhysRevA.90.043611}.

\bibitem[{\citenamefont{{Z}wierlein et~al.}(2005)\citenamefont{{Z}wierlein,
  {A}bo {S}haeer, {S}chirotzek, {S}chunck, and {K}etterle}}]{Zwierlein2005}
\bibinfo{author}{\bibfnamefont{M.~W.} \bibnamefont{{Z}wierlein}},
  \bibinfo{author}{\bibfnamefont{J.~R.} \bibnamefont{{A}bo {S}haeer}},
  \bibinfo{author}{\bibfnamefont{A.}~\bibnamefont{{S}chirotzek}},
  \bibinfo{author}{\bibfnamefont{C.~H.} \bibnamefont{{S}chunck}},
  \bibnamefont{and}
  \bibinfo{author}{\bibfnamefont{W.}~\bibnamefont{{K}etterle}},
  \bibinfo{journal}{{N}ature} \textbf{\bibinfo{volume}{435}},
  \bibinfo{pages}{1047} (\bibinfo{year}{2005}),
  \urlprefix\url{http://dx.doi.org/10.1038/nature03858}.

\bibitem[{\citenamefont{Plisson et~al.}(2011)\citenamefont{Plisson, Allard,
  Holzmann, Salomon, Aspect, Bouyer, and Bourdel}}]{Plisson2011}
\bibinfo{author}{\bibfnamefont{T.}~\bibnamefont{Plisson}},
  \bibinfo{author}{\bibfnamefont{B.}~\bibnamefont{Allard}},
  \bibinfo{author}{\bibfnamefont{M.}~\bibnamefont{Holzmann}},
  \bibinfo{author}{\bibfnamefont{G.}~\bibnamefont{Salomon}},
  \bibinfo{author}{\bibfnamefont{A.}~\bibnamefont{Aspect}},
  \bibinfo{author}{\bibfnamefont{P.}~\bibnamefont{Bouyer}}, \bibnamefont{and}
  \bibinfo{author}{\bibfnamefont{T.}~\bibnamefont{Bourdel}},
  \bibinfo{journal}{Phys. Rev. A} \textbf{\bibinfo{volume}{84}},
  \bibinfo{pages}{061606} (\bibinfo{year}{2011}),
  \urlprefix\url{http://link.aps.org/doi/10.1103/PhysRevA.84.061606}.

\bibitem[{\citenamefont{Prokof'ev et~al.}(2001)\citenamefont{Prokof'ev,
  Ruebenacker, and Svistunov}}]{Prokofev2001}
\bibinfo{author}{\bibfnamefont{N.}~\bibnamefont{Prokof'ev}},
  \bibinfo{author}{\bibfnamefont{O.}~\bibnamefont{Ruebenacker}},
  \bibnamefont{and}
  \bibinfo{author}{\bibfnamefont{B.}~\bibnamefont{Svistunov}},
  \bibinfo{journal}{Phys. Rev. Lett.} \textbf{\bibinfo{volume}{87}},
  \bibinfo{pages}{270402} (\bibinfo{year}{2001}),
  \urlprefix\url{http://link.aps.org/doi/10.1103/PhysRevLett.87.270402}.

\bibitem[{\citenamefont{Prokof'ev and Svistunov}(2002)}]{Prokofev2002}
\bibinfo{author}{\bibfnamefont{N.}~\bibnamefont{Prokof'ev}} \bibnamefont{and}
  \bibinfo{author}{\bibfnamefont{B.}~\bibnamefont{Svistunov}},
  \bibinfo{journal}{Phys. Rev. A} \textbf{\bibinfo{volume}{66}},
  \bibinfo{pages}{043608} (\bibinfo{year}{2002}),
  \urlprefix\url{http://link.aps.org/doi/10.1103/PhysRevA.66.043608}.

\bibitem[{\citenamefont{Bisset et~al.}(2009)\citenamefont{Bisset, Davis,
  Simula, and Blakie}}]{Bisset2009}
\bibinfo{author}{\bibfnamefont{R.~N.} \bibnamefont{Bisset}},
  \bibinfo{author}{\bibfnamefont{M.~J.} \bibnamefont{Davis}},
  \bibinfo{author}{\bibfnamefont{T.~P.} \bibnamefont{Simula}},
  \bibnamefont{and} \bibinfo{author}{\bibfnamefont{P.~B.}
  \bibnamefont{Blakie}}, \bibinfo{journal}{Phys. Rev. A}
  \textbf{\bibinfo{volume}{79}}, \bibinfo{pages}{033626}
  (\bibinfo{year}{2009}),
  \urlprefix\url{http://link.aps.org/doi/10.1103/PhysRevA.79.033626}.

\bibitem[{\citenamefont{Clad\'e et~al.}(2009)\citenamefont{Clad\'e, Ryu,
  Ramanathan, Helmerson, and Phillips}}]{Clade2009}
\bibinfo{author}{\bibfnamefont{P.}~\bibnamefont{Clad\'e}},
  \bibinfo{author}{\bibfnamefont{C.}~\bibnamefont{Ryu}},
  \bibinfo{author}{\bibfnamefont{A.}~\bibnamefont{Ramanathan}},
  \bibinfo{author}{\bibfnamefont{K.}~\bibnamefont{Helmerson}},
  \bibnamefont{and} \bibinfo{author}{\bibfnamefont{W.~D.}
  \bibnamefont{Phillips}}, \bibinfo{journal}{Phys. Rev. Lett.}
  \textbf{\bibinfo{volume}{102}}, \bibinfo{pages}{170401}
  (\bibinfo{year}{2009}),
  \urlprefix\url{http://link.aps.org/doi/10.1103/PhysRevLett.102.170401}.

\bibitem[{\citenamefont{Hung et~al.}(2011)\citenamefont{Hung, Zhang, Gemelke,
  and Chin}}]{Hung2011a}
\bibinfo{author}{\bibfnamefont{C.-L.} \bibnamefont{Hung}},
  \bibinfo{author}{\bibfnamefont{X.}~\bibnamefont{Zhang}},
  \bibinfo{author}{\bibfnamefont{N.}~\bibnamefont{Gemelke}}, \bibnamefont{and}
  \bibinfo{author}{\bibfnamefont{C.}~\bibnamefont{Chin}},
  \bibinfo{journal}{Nature} \textbf{\bibinfo{volume}{470}},
  \bibinfo{pages}{236} (\bibinfo{year}{2011}),
  \urlprefix\url{http://dx.doi.org/10.1038/nature09722}.

\bibitem[{\citenamefont{Petrov et~al.}(2003)\citenamefont{Petrov, Baranov, and
  Shlyapnikov}}]{Petrov2003}
\bibinfo{author}{\bibfnamefont{D.~S.} \bibnamefont{Petrov}},
  \bibinfo{author}{\bibfnamefont{M.~A.} \bibnamefont{Baranov}},
  \bibnamefont{and} \bibinfo{author}{\bibfnamefont{G.~V.}
  \bibnamefont{Shlyapnikov}}, \bibinfo{journal}{Phys. Rev. A}
  \textbf{\bibinfo{volume}{67}}, \bibinfo{pages}{031601}
  (\bibinfo{year}{2003}),
  \urlprefix\url{http://link.aps.org/doi/10.1103/PhysRevA.67.031601}.

\bibitem[{\citenamefont{Ngampruetikorn
  et~al.}(2013)\citenamefont{Ngampruetikorn, Levinsen, and
  Parish}}]{Ngampruetikorn2013}
\bibinfo{author}{\bibfnamefont{V.}~\bibnamefont{Ngampruetikorn}},
  \bibinfo{author}{\bibfnamefont{J.}~\bibnamefont{Levinsen}}, \bibnamefont{and}
  \bibinfo{author}{\bibfnamefont{M.~M.} \bibnamefont{Parish}},
  \bibinfo{journal}{Phys. Rev. Lett.} \textbf{\bibinfo{volume}{111}},
  \bibinfo{pages}{265301} (\bibinfo{year}{2013}),
  \urlprefix\url{http://link.aps.org/doi/10.1103/PhysRevLett.111.265301}.

\bibitem[{\citenamefont{S\'a~de Melo et~al.}(1993)\citenamefont{S\'a~de Melo,
  Randeria, and Engelbrecht}}]{SadeMelo1993}
\bibinfo{author}{\bibfnamefont{C.~A.~R.} \bibnamefont{S\'a~de Melo}},
  \bibinfo{author}{\bibfnamefont{M.}~\bibnamefont{Randeria}}, \bibnamefont{and}
  \bibinfo{author}{\bibfnamefont{J.~R.} \bibnamefont{Engelbrecht}},
  \bibinfo{journal}{Phys. Rev. Lett.} \textbf{\bibinfo{volume}{71}},
  \bibinfo{pages}{3202} (\bibinfo{year}{1993}),
  \urlprefix\url{http://link.aps.org/doi/10.1103/PhysRevLett.71.3202}.

\bibitem[{\citenamefont{{Fischer} and {Parish}}(2014)}]{Fischer2014}
\bibinfo{author}{\bibfnamefont{A.~M.} \bibnamefont{{Fischer}}}
  \bibnamefont{and} \bibinfo{author}{\bibfnamefont{M.~M.}
  \bibnamefont{{Parish}}}, \bibinfo{journal}{ArXiv e-prints}
  (\bibinfo{year}{2014}), \eprint{1408.0476},
  \urlprefix\url{http://arxiv.org/abs/1408.0476}.

\bibitem[{\citenamefont{Dyke et~al.}(2014)\citenamefont{Dyke, Fenech, Peppler,
  Lingham, Hoinka, Zhang, Mulkerin, Hu, Liu, and Vale}}]{Dyke2014}
\bibinfo{author}{\bibfnamefont{P.}~\bibnamefont{Dyke}},
  \bibinfo{author}{\bibfnamefont{K.}~\bibnamefont{Fenech}},
  \bibinfo{author}{\bibfnamefont{T.}~\bibnamefont{Peppler}},
  \bibinfo{author}{\bibfnamefont{M.~G.} \bibnamefont{Lingham}},
  \bibinfo{author}{\bibfnamefont{S.}~\bibnamefont{Hoinka}},
  \bibinfo{author}{\bibfnamefont{W.}~\bibnamefont{Zhang}},
  \bibinfo{author}{\bibfnamefont{B.}~\bibnamefont{Mulkerin}},
  \bibinfo{author}{\bibfnamefont{H.}~\bibnamefont{Hu}},
  \bibinfo{author}{\bibfnamefont{X.~J.} \bibnamefont{Liu}}, \bibnamefont{and}
  \bibinfo{author}{\bibfnamefont{C.~J.} \bibnamefont{Vale}}
  (\bibinfo{year}{2014}), \eprint{1411.4703}.

\bibitem[{\citenamefont{Tan}(2008)}]{Tan2008}
\bibinfo{author}{\bibfnamefont{S.}~\bibnamefont{Tan}}, \bibinfo{journal}{Annals
  of Physics} \textbf{\bibinfo{volume}{323}}, \bibinfo{pages}{2971 }
  (\bibinfo{year}{2008}), ISSN \bibinfo{issn}{0003-4916},
  \urlprefix\url{http://www.sciencedirect.com/science/article/pii/S0003491608000432}.

\bibitem[{\citenamefont{{P}etrov and {S}hlyapnikov}(2001)}]{Petrov2001}
\bibinfo{author}{\bibfnamefont{D.~S.} \bibnamefont{{P}etrov}} \bibnamefont{and}
  \bibinfo{author}{\bibfnamefont{G.~V.} \bibnamefont{{S}hlyapnikov}},
  \bibinfo{journal}{{P}hys. {R}ev. {A}} \textbf{\bibinfo{volume}{64}},
  \bibinfo{pages}{012706} (\bibinfo{year}{2001}),
  \urlprefix\url{http://link.aps.org/abstract/PRA/v64/e012706}.

\bibitem[{\citenamefont{Petrov et~al.}(2005)\citenamefont{Petrov, Salomon, and
  Shlyapnikov}}]{Petrov2005}
\bibinfo{author}{\bibfnamefont{D.~S.} \bibnamefont{Petrov}},
  \bibinfo{author}{\bibfnamefont{C.}~\bibnamefont{Salomon}}, \bibnamefont{and}
  \bibinfo{author}{\bibfnamefont{G.~V.} \bibnamefont{Shlyapnikov}},
  \bibinfo{journal}{Phys. Rev. A} \textbf{\bibinfo{volume}{71}},
  \bibinfo{pages}{012708} (\bibinfo{year}{2005}),
  \urlprefix\url{http://link.aps.org/doi/10.1103/PhysRevA.71.012708}.

\bibitem[{\citenamefont{Fr{\"o}hlich et~al.}(2011)\citenamefont{Fr{\"o}hlich,
  Feld, Vogt, Koschorreck, Zwerger, and K\"ohl}}]{froehlich2011}
\bibinfo{author}{\bibfnamefont{B.}~\bibnamefont{Fr{\"o}hlich}},
  \bibinfo{author}{\bibfnamefont{M.}~\bibnamefont{Feld}},
  \bibinfo{author}{\bibfnamefont{E.}~\bibnamefont{Vogt}},
  \bibinfo{author}{\bibfnamefont{M.}~\bibnamefont{Koschorreck}},
  \bibinfo{author}{\bibfnamefont{W.}~\bibnamefont{Zwerger}}, \bibnamefont{and}
  \bibinfo{author}{\bibfnamefont{M.}~\bibnamefont{K\"ohl}},
  \bibinfo{journal}{Phys. Rev. Lett.} \textbf{\bibinfo{volume}{106}},
  \bibinfo{pages}{105301} (\bibinfo{year}{2011}),
  \urlprefix\url{http://link.aps.org/doi/10.1103/PhysRevLett.106.105301}.

\bibitem[{\citenamefont{Reinaudi et~al.}(2007)\citenamefont{Reinaudi, Lahaye,
  Wang, and Gu\'{e}ry-Odelin}}]{Reinaudi2007}
\bibinfo{author}{\bibfnamefont{G.}~\bibnamefont{Reinaudi}},
  \bibinfo{author}{\bibfnamefont{T.}~\bibnamefont{Lahaye}},
  \bibinfo{author}{\bibfnamefont{Z.}~\bibnamefont{Wang}}, \bibnamefont{and}
  \bibinfo{author}{\bibfnamefont{D.}~\bibnamefont{Gu\'{e}ry-Odelin}},
  \bibinfo{journal}{Opt. Lett.} \textbf{\bibinfo{volume}{32}},
  \bibinfo{pages}{3143} (\bibinfo{year}{2007}),
  \urlprefix\url{http://ol.osa.org/abstract.cfm?URI=ol-32-21-3143}.

\bibitem[{\citenamefont{Yefsah et~al.}(2011)\citenamefont{Yefsah, Desbuquois,
  Chomaz, G\"unter, and Dalibard}}]{Yefsah2011}
\bibinfo{author}{\bibfnamefont{T.}~\bibnamefont{Yefsah}},
  \bibinfo{author}{\bibfnamefont{R.}~\bibnamefont{Desbuquois}},
  \bibinfo{author}{\bibfnamefont{L.}~\bibnamefont{Chomaz}},
  \bibinfo{author}{\bibfnamefont{K.~J.} \bibnamefont{G\"unter}},
  \bibnamefont{and} \bibinfo{author}{\bibfnamefont{J.}~\bibnamefont{Dalibard}},
  \bibinfo{journal}{Phys. Rev. Lett.} \textbf{\bibinfo{volume}{107}},
  \bibinfo{pages}{130401} (\bibinfo{year}{2011}),
  \urlprefix\url{http://link.aps.org/doi/10.1103/PhysRevLett.107.130401}.

\bibitem[{\citenamefont{Gehm}(2003)}]{gehm2003}
\bibinfo{author}{\bibfnamefont{M.~E.} \bibnamefont{Gehm}},
  \bibinfo{journal}{Properties of $^6$Li}  (\bibinfo{year}{2003}),
  \urlprefix\url{www.physics.ncsu.edu/jet/techdocs/pdf/PropertiesOfLi.pdf}.

\bibitem[{\citenamefont{Ockeloen et~al.}(2010)\citenamefont{Ockeloen,
  Tauschinsky, Spreeuw, and Whitlock}}]{Ockeloen2010}
\bibinfo{author}{\bibfnamefont{C.~F.} \bibnamefont{Ockeloen}},
  \bibinfo{author}{\bibfnamefont{A.~F.} \bibnamefont{Tauschinsky}},
  \bibinfo{author}{\bibfnamefont{R.~J.~C.} \bibnamefont{Spreeuw}},
  \bibnamefont{and} \bibinfo{author}{\bibfnamefont{S.}~\bibnamefont{Whitlock}},
  \bibinfo{journal}{Phys. Rev. A} \textbf{\bibinfo{volume}{82}},
  \bibinfo{pages}{061606} (\bibinfo{year}{2010}),
  \urlprefix\url{http://link.aps.org/doi/10.1103/PhysRevA.82.061606}.

\bibitem[{\citenamefont{Ketterle and Zwierlein}(2008)}]{Ketterle2008}
\bibinfo{author}{\bibfnamefont{W.}~\bibnamefont{Ketterle}} \bibnamefont{and}
  \bibinfo{author}{\bibfnamefont{M.~W.} \bibnamefont{Zwierlein}},
  \bibinfo{journal}{Rivista del nuovo cimento} \textbf{\bibinfo{volume}{5-6}},
  \bibinfo{pages}{247} (\bibinfo{year}{2008}),
  \urlprefix\url{http://dx.doi.org/10.1393/ncr/i2008-10033-1}.

\end{thebibliography}
\end{document}